\documentclass[12pt,a4paper]{article}

\usepackage[british]{babel}

\usepackage[a4paper,top=2cm,bottom=2cm,left=3cm,right=3cm,marginparwidth=2.5cm]{geometry}

\usepackage{amsmath}
\usepackage{graphicx}
\usepackage[colorlinks=true, allcolors=blue]{hyperref}
\usepackage{url}
\usepackage[title]{appendix}
\usepackage{mathrsfs}
\usepackage{amsfonts}
\usepackage{booktabs} 
\usepackage{caption}  
\usepackage{threeparttable} 
\usepackage{algorithm}
\usepackage{algorithmicx}
\usepackage{algpseudocode}
\usepackage{listings}
\usepackage{enumitem}
\usepackage{chngcntr}
\usepackage{booktabs}
\usepackage{lipsum}
\usepackage{subcaption}
\usepackage[T1]{fontenc}    
\usepackage{csquotes}       
\usepackage{diagbox}
\usepackage{authblk}

\usepackage{setspace}
\title{Enabling Secure and Ephemeral AI Workloads \\in Data Mesh Environments}
\author[1]{Chinkit Patel\footnote{Primary contact: chinkit.patel@gmail.com}}
\author[2]{Kee Siong Ng}

\affil[1]{The Enigma Co Pty Ltd}
\affil[2]{The Australian National University}

\date{}  

\begin{document}
\maketitle

\begin{abstract}
Many large enterprises that operate highly governed and complex ICT environments have no efficient and effective way to support their Data and AI teams in rapidly spinning up and tearing down self-service data and compute infrastructure, to experiment with new data analytic tools, and deploy data products into operational use \cite{donoho201750, kim2017data}.
This paper proposes a key piece of the solution to the overall problem, in the form of an on-demand self-service data-platform infrastructure to empower de-centralised data teams to build data products on top of centralised templates, policies and governance.
The core innovation is an efficient method to leverage immutable container operating systems \cite{bohm2023immutable} and infrastructure-as-code methodologies for creating, from scratch, vendor-neutral and short-lived Kubernetes clusters on-premises and in any cloud environment.
Our proposed approach can serve as a repeatable, portable and cost-efficient alternative or complement to commercial Platform as a Service (PaaS) offerings, and this is particularly important in supporting interoperability in complex data mesh environments \cite{dehghani2022data} with a mix of modern and legacy compute infrastructure.
\end{abstract}


\newpage
\tableofcontents
\newpage


\section{Introduction}\label{sec:intro}

Data mesh \cite{dehghani2022data} is a decentralised data architecture and organisational-design approach that has been introduced recently to overcome the limitations of traditional centralised data systems like data warehouses and data lakes. 
It emphasises domain-oriented ownership, treating data as a product, and enabling self-service capabilities, all governed by federated computational standards. 
Since its introduction in 2019, the data mesh concept has gained significant traction among organisations that seek to scale their data and AI operations effectively.

A key component of a data mesh architecture is the availability of a self-serve data and compute infrastructure platform.
This can be a challenging infrastructure to build, given the need to carefully balance innovation, security, compliance, and trust. 
Indeed, despite significant investments in modern DevSecOps practices and cloud platforms, many large organisations that operate highly governed and complex ICT environments still struggle to provide infrastructure that can effectively support their Data and AI teams with basic requirements like being able to 
\begin{enumerate}\itemsep0mm\parskip0mm
    \item rapidly spin up and tear down self-serve data and compute infrastructure for different types of AI and data workloads; 
    \item easily experiment with new data and AI tools to find ones that are fit for purpose for different use cases; and 
    \item efficiently deploy data and AI products into operational use and monitor their performance, especially for restricted ICT environments like edge devices. 
\end{enumerate}
These basic infrastructural challenges prevent organisations from being able to fully realise the benefits of becoming data-driven and can be a major impediment to staff productivity and organisational efficiency.

This paper introduces an architectural pattern to build on-demand, self-service AI and data engineering platforms in decentralised data mesh architectures. 
The architecture pattern provides balanced autonomy, allowing developers to efficiently and effectively experiment with new data analytics tools and techniques in development and testing environments, 
and then securely promote their apps for deployment in secure, standardised environments. 
The core innovation is an efficient method to leverage immutable container operating systems \cite{bohm2023immutable} and infrastructure-as-code methodologies for creating, from scratch, vendor-neutral and short-lived Kubernetes clusters on-premises and in any cloud environment.
Our proposed approach can serve as a repeatable, portable and cost-efficient alternative or complement to commercial Platform as a Service (PaaS) offerings, and this is particularly important in supporting interoperability in complex data mesh environments \cite{dehghani2022data} with a mix of modern and legacy compute infrastructure.
The architecture pattern is implemented in a prototype called \texttt{sskuba} (pronounced `scuba' and stands for Self-service Kubernetes for AI).

\section{Barriers to AI Innovation in Large Enterprises}\label{sec:barriers}

By some estimates \cite{donoho201750, kim2017data, Hurtgen2020}, Data and AI teams are spending 50-80\% of their time seeking and managing compute infrastructure rather than focusing on analytics and insights.  
A national survey in Australia\footnote{\href{https://www.csiro.au/-/media/D61/Files/22-00724_DATA61_INFOGRAPHIC_AccessAI.pdf}{Industry access to AI computing infrastructure and services by Australia's National AI Centre}} 
shows upwards of 70\% of AI scientists and engineers are concerned about two issues: (i) compute affordability and availability, and (ii) platform and data security. The latter is in part because of the complex interplay between AI systems and cyber and data security \cite{roy2023survey, habbal2024artificial}. 
This section points out some of the reasons why it is hard to innovate and adopt modern AI and data technologies in large traditional enterprises, and how the emerging architecture concept of Data Mesh, complemented by the \texttt{sskuba} platform, can be used to address some of the challenges.

\subsection{Providing Self-Serve AI Compute Infrastructure is Hard}\label{subsec:k8s is hard}

The success of AWS within Amazon \cite{vogels2006learning}, and that of Borg, Omega and Kubernetes within Google \cite{verma2015large, burns2016borg}, followed by the subsequent industry-wide adoption of these cloud technologies, have spawned many a platform teams in organisations large and small.
The basic idea is that sustainable agility can only be achieved by first creating platforms that are reliable, repeatable and agile \cite{skelton2019team}.

However, in large regulated organisations, platform teams are toiling with the plethora of cloud technologies from different vendors.  
Platform teams also often face challenges to influence organisational governance processes and decision points, and have to fight for their survival in project-based funding models. 
For those that survive, they usually accumulate high tacit knowledge, lack "bus factor" coverage and have a tendency to address complexity through conformity, compliance, and centralisation.
These issues often manifest themselves in two barriers to innovation:
\begin{itemize}\itemsep0mm\parskip0mm
    \item insecure use of cloud, as evident by the fact that misconfiguration remains the number one issue in cloud security \cite{CSA2024}; and
    \item ineffective and frustrated data and AI teams who find themselves spending significant amount of time struggling with basic compute infrastructure and ICT integration issues rather than doing data analytics and AI modelling.
\end{itemize}

Industry standardisation on Linux containers and Kubernetes certainly help, but they are not the full answers.
For background, containers are Linux processes that are isolated using three Linux kernel mechanisms: (i) namespaces to limit what the container process can see, for example by giving the container an isolated set of process IDs; (ii) cgroups to control what resources and capabilities the container process can access; and (iii) chroot to limit the set of files and directories the container process can see \cite{rice2020container}.
Containers are useful for packaging and running images as isolated workloads on machines.
Kubernetes is an orchestrator of containerised cloud-native microservices apps. 
It is used to deploy and manage applications that are packaged as containers and provide easy scaling, update and self-heal operations for these applications \cite{poulton2024kubernetes}.
A key component of Kubernetes is the reconciliation loop. There are three states of the world: (i) a desired state, which is a declarative statement of what the world should be like; (ii) an actual state, which can be quite complex; and (iii) an observed state, which gives the observable / measurable parts of the actual state and may be noisy, incomplete, or out of date. 
The role of the Kubernetes reconciliation loop is to repeatedly compare the current observed state against the desired state, and then take action to change the actual state to match the desired state. 
The reconciliation control loop is what transforms Kubernetes into a self-healing, dynamic system by automatically causing it to restore the system to the desired state without needing operator intervention \cite{burns15}. 

\paragraph{Setting up Kubernetes clusters is hard}
Whilst Kubernetes has become the de facto container orchestration platform with good developer and user experience, major annual surveys like The Kubernetes Benchmark Report by Fairwinds consistently show that many organisations continue to struggle with the security and integration aspects of setting up and managing Kubernetes clusters. The following are some of the most prominent challenges.
\begin{itemize}
    \item \textbf{Installation} - The possible configuration space for deployments is immense \cite{bryant2024kubernetes}. 
    Managed Kubernetes PaaS vendors provide configuration managers (e.g. \texttt{eksctl}) however they are not vendor-neutral and usually do not deal with multi-cloud deployments. Popular open-source installers like Cluster API and Kubespray are difficult to use without an intricate understanding of the inner workings of Kubernetes and its components. 
    Some rely on creating a management Kubernetes cluster, thus creating a chicken-and-egg problem in secure standardised environments. 
    Others, like Kubitect \cite{muvsic2024digital}, use a two-step installation process, whereby a base Operating System (OS) is first installed, and then the nodes are further prepared by installing Kubernetes components with Ansible or Terraform. This approach contradicts the principle of cloud-native immutable infrastructure and can lead to fragile installation processes and mutable systems that evolve over time in unknowable ways \cite{mikkelsen2019immutable}. 
    
    \item \textbf{Multi-Tenancy} - Kubernetes lacks hard multi-tenancy support like virtual machines, with namespaces providing only logical isolation. The vulnerability of such weak boundary control was exposed in Aug 2024, when Mandiant reported an Azure Kubernetes Service exploit that allows an attacker to privilege-escalate and read all the secrets within a cluster\footnote{\url{https://msrc.microsoft.com/update-guide/vulnerability/CVE-2024-21403}}. Furthermore, most tools deployed on Kubernetes use the operator pattern, which means changes and upgrades require coordinated testing across all tenants because Kubernetes operators, while namespace scoped, often require cluster-wide resources and permissions.  
    This makes centralised clusters impractical for catering to the diverse needs of hundreds of developers. 
    

    \item \textbf{Integration with Enterprise ICT Services} - 
    Kubernetes clusters need to be well integrated with enterprise services such as single sign-on, container hardening pipeline, public key infrastructure and logging and SIEM monitoring. 
    This is particularly challenging in complex hybrid cloud environments with multiple security domains, where highly customised integration work is needed for each platform that hosts a Kubernetes cluster.   

\end{itemize}

To compound these challenges, Kubernetes has a vast ecosystem that can be hard to navigate.
At the time of writing, there are 60 Kubernetes distributions, 59 hosted solutions, and 19 installers that have been certified by the Cloud Native Computing Foundation (CNCF), as shown in Appendix~\ref{app:k8s distributions}. 
The complexity of setting up and managing Kubernetes clusters is the reason why many organisations use managed Kubernetes services from cloud vendors. 
While this is a reasonable position to take for some organisations, it introduces risks like vendor lock-in and interoperability issues in complex, distributed data-mesh environments. 
Our proposed solution, and a comparison with key alternatives, are described in \S\ref{sec:self-service k8s}.

\subsection{Getting Data is Hard}

\begin{figure}
    \centering
    \includegraphics[width=0.95\textwidth]{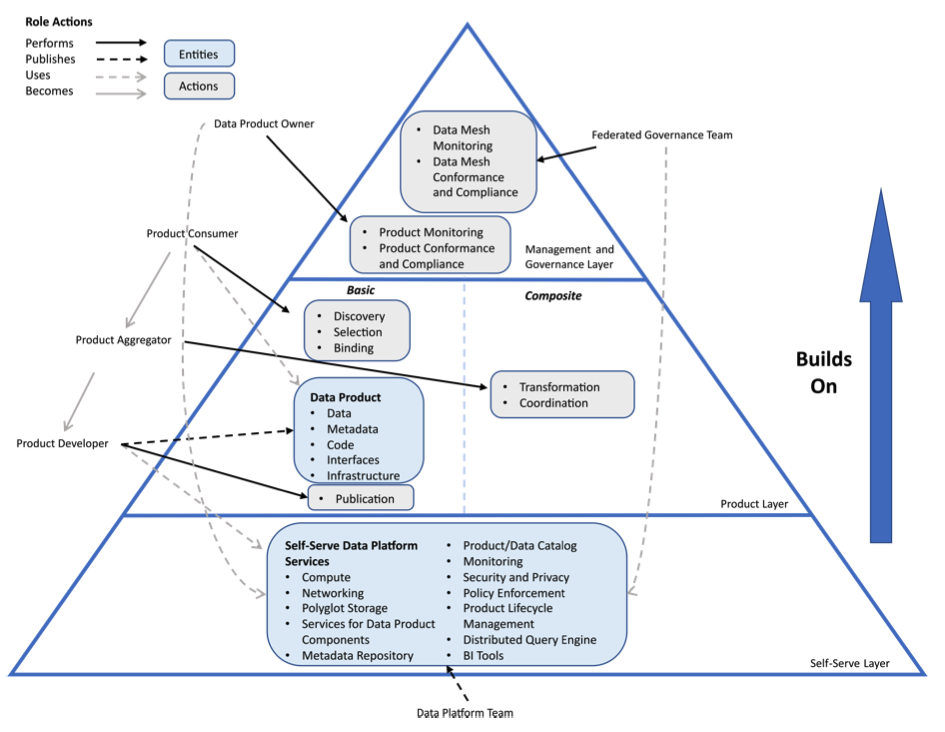}
    \caption{Layered architecture for data mesh (source: \cite{goedegebuure2024data})}
    \label{fig:data mesh}
\end{figure}

A common issue for data scientists and AI engineers in large, complex organisations is in getting access to data in the first place \cite{coleman2016can}.
This can occur because the underlying ICT and data infrastructure may be old and not sufficiently integrated, resulting in siloed data sources that also often have poor data quality.
There could also be cultural issues preventing sharing of data between organisational units.
Consolidating data into a central repository, such as a data lake or enterprise data warehouse, is a common strategy for breaking down silos and make data more widely accessible \cite{nambiar2022overview}.
Unfortunately, this approach cannot be easily scaled to large organisations with complex data ownership structures and federated ICT environments.
The failure modes of monolithic and centralised data warehouses and lakes are studied in  \cite{dehghani2022data} and a new data mesh architecture was proposed. 
The data mesh architecture significantly enhances access to data and the building of data products through several key principles and features:
\begin{description}
    \item [Decentralized Data Ownership]
     Data mesh architecture decentralises data ownership, assigning responsibility for data management to the domain teams that generate and use the data. This approach ensures that data is managed by those who best understand it, reducing reliance on a central IT team and improving data accessibility and quality.

    \item [Data as a Product]
     Treating data as a product is a core principle of data mesh. Domain teams are encouraged to think of their data as products that need to meet the needs of other domains within the organisation. This product thinking ensures that data is made discoverable, usable, and of high quality, similar to how a product would be developed for external customers.
     Data products are registered in a centralised data catalog, making it easily discoverable by others. This approach enhances data accessibility, reduces the time it takes to obtain data, and improves business agility.

    \item [Self-Serve Data Infrastructure]
     A self-serve infrastructure enables domain teams to create and manage their own data products without having to rely on constant support from a central IT team. Such a self-serve infrastructure would provide tools and technologies to streamline processes such as data ingestion, transformation, and storage while reducing duplicated efforts and bottlenecks.

    \item [Federated Computational Governance]
     While each domain has autonomy under the decentralised data ownership model, federated computational governance ensures there is a centralised governance framework that sets standards, policies, and processes across all domains. This governance model promotes consistency, interoperability, and compliance, making it easier for data to be shared and used across different business units, thus facilitating cross-team collaboration.

\end{description}     
The data mesh architecture is being increasingly adopted in large organisations, but there are many pitfalls associated with its implementation \cite{goedegebuure2024data, bode2024towards}.
Figure~\ref{fig:data mesh} shows a layered model of the data mesh. 
The separation into layers highlights the different needs regarding the provision of infrastructure for
i) creating and composing data products of differing complexities; 
ii) serving data products to consumers securely and at scale;  
and iii) managing data products throughout their life cycles.
In organisations with complex ICT environments, the self-serve data-platform services that underpins the success of the overall data mesh architecture is often inadequate or missing altogether.
The \texttt{sskuba} solution architecture proposed in this paper is designed explicitly to address this need.

\subsection{Deploying AI Solutions to Edge Devices is Hard}

In the military and intelligence context, the concept of software-defined warfare \cite{mulchandani2022software} is gaining mind share. A central thesis of software-defined warfare (SDW) is that in future crises and conflicts, the side that can adapt faster and demonstrate the greatest agility in rapidly updating and promulgating fielded software and AI models is likely to gain significant decision advantage.
A key enabler for SDW is a collaborative and secure DevOps environment for AI scientists, software developers and operational users to work closely together at a high tempo to produce, deploy and evaluate innovative solutions in limited operations (LimOps) that are constrained in scope, duration, scale, and/or resources but can play a crucial role in providing targeted insights and actions that align with broader strategic goals. 
Key applications include intelligence gathering (e.g. the temporary instrumentation of a data collection and analysis capability),
counter-terrorism efforts (e.g. targeted deployment of offensive cyber capabilities),
and humanitarian assistance (e.g. quick setup of communication infrastructure in resource-constrained scenarios). 
Given LimOps may be time-sensitive and require rapid execution with only limited resources, it is natural for LimOps to be conducted on temporary extensions of an existing data mesh, and such extensions would usually be sufficiently isolated from the main data mesh network and be afforded a lower bar for product deployment.
This same scenario applies to other disciplines where there is a need to rapidly develop and deploy AI solutions under heavily constrained ICT environments, including mining, agriculture, environmental monitoring, and other industrial internet-of-things applications.

The \texttt{sskuba} platform is designed to support ephemeral AI workload in a Data Mesh architecture using minimal resources.
From that perspective, it has a niche application in providing that collaborative and secure LimOps environment.
There are alternatives in the form of lightweight Kubernetes, of course, and a comparison of Kubernetes distributions for the edge can be found in \cite{kjorveziroski2022kubernetes}.

\subsection{Choosing Modern AI and Data Tool Kits is Hard}

It seems not a day goes by without yet another data and/or AI system being invented these days. 
Indeed, the number and variety of data and AI tools available in the marketplace have exploded in the last 10-15 years, as can be seen in the two snapshots given in Appendix~\ref{app:data AI landscape}. 
Further, we can expect the number of programming languages, systems and platforms for AI to continue to grow and the useful half-life of each system to get shorter and shorter. 
This is all well and good for scientific and societal progress, but this Cambrian explosion in inventions and activities can be overwhelming and confusing to AI scientists, platform engineers, and enterprise architects faced with the paradox of choice in having to decide which AI and data tool kits to adopt within an enterprise.

The above phenomenon has important implications for organisations, and our hypothesis is that those organisations that commit to one single platform with a heavy investment will likely suffer from loss of agility over time (perhaps as quickly as 1-2 years), and those that invest in adaptability and resilience, by skilling up its people with versatile tool kits and the essential knowledge to work with many different ways of representing data and performing operations on them, will be the ones that survive and thrive.

The \texttt{sskuba} platform is designed to provide organisations with an agile and safe AI and compute infrastructure to evaluate and adopt the diverse set of analytics tools getting invented everyday, with a modular architecture that affords flexibility in swapping in and out component technologies.

\section{The Role of \texttt{sskuba} in a DevSecOps Framework}\label{sec:broader ecosystem}

Having explained how \texttt{sskuba} can help address the major barriers to AI adoption in large organisations as part of a Data Mesh architecture in \S\ref{sec:barriers}, we now seek to explain how \texttt{sskuba} fits in with modern software engineering practices, in particular its role in a modern DevSecOps framework.

We first note that the US Department of Defense (DOD) has published a mature set of guidance and policy documents for building modern software, including a DevSecOps Playbook, reference designs for setting up and maintaining Kubernetes clusters, guidance on operating software factories and use of open-source software, and a Continuous Authority to Operate (cATO) implementation guide.
These reference documents can be found at \url{https://dodcio.defense.gov/library/} and we recommend them as a strong starting point for organisations of all sizes.

Another industry trend worth noting is that there are now many mature and sophisticated data analytics and AI platforms and products that come ready to be installed on a Kubernetes cluster. For example, Platform One (P1) Big Bang by U.S. Air Force, which allows organisations to build their own DevSecOps platform using hardened and approved packages, and Anduril's Lattice Mesh, which can be used to do sophisticated data integration and data fusion across complex ICT environments, and many other products all adopt this bring-your-own-kubernetes (BYK8s) deployment process.

Building on the DOD and general industry best practices like BYK8s, there are essentially four layers of infrastructure required to rapidly develop and deploy AI apps in a data mesh environment, supported by an overall DevSecOps playbook: 
\begin{enumerate}\itemsep1mm\parskip0mm
    \item A self-service (enterprise-grade) Kubernetes infrastructure in the cloud or on-premises with baked-in security; 
    \item A container-hardening pipeline and a repository of hardened artifact repository (e.g. Iron Bank and ChainGuard);
    \item A software factory that leverages infrastructure automation, hardened containers, development tools, and continuous integration / continuous deployment (CI/CD) pipelines to build and deploy resilient software at speed in a way that minimises software supply chain risks.
    
    \item The equivalent of a Continuous Authority to Operate process, whereby system owners must be able to demonstrate (i) continuous monitoring of cybersecurity controls and risks; and (ii) ability to conduct active cyber defense in (near) real time. 
\end{enumerate}

The \texttt{sskuba} platform is primarily a contribution to a self-service Kubernetes infrastructure for AI workloads. 
It works within an overall DevSecOps playbook and uses container-hardening pipeline and the software factory to build and deploy AI and data applications.
The secondary contribution of \texttt{sskuba} is the provision of a curated set of mature, open-source AI and data tool kits that can help AI solution developers and data engineers hit the ground running from day one.

The design of the self-service Kubernetes component of \texttt{sskuba} is described in \S\ref{sec:self-service k8s}; the design of the Data and AI tool kits component is described in \S\ref{sec:AI and data tools}.

\section{Self-Service Kubernetes Anywhere}\label{sec:self-service k8s}

We first describe the principles and core technologies that underlie the self-service Kubernetes architecture of \texttt{sskuba}, whose design goal is to enable hundreds of de-centralised data teams to reliably and securely create cost efficient and trusted data products at the speed of relevance. 
This is followed by a detailed description of how \texttt{sskuba} clusters are built. Alternative technologies are discussed at the end. 

\subsection{Design Principles}\label{subsec:principles} 
Here are the principles and best practices that guide the design of the self-service Kubernetes cluster setup component of \texttt{sskuba}.

\begin{enumerate}
    \item \textbf{Easy to deploy anywhere} \label{subsubsec:deploy anywhere} \\
    Hybrid-cloud and multi-cloud architectures offer numerous advantages but also introduce significant challenges and complexities \cite{Seth2024Navigating, hong2019overview, ferry2018cloudmf}. 
    Kubernetes clusters should be simple and cost-effective to setup and deploy across hyperscaler clouds, on-premises, edge, and hybrid environments. To achieve this, all data and AI tools must support self-hosted mode, with packaging for libraries, containers, and repositories to enable full functionality in disconnected environments. This principle ensures portability, leverages cloud innovation, rides the hardware commoditisation curve, and helps with compliance with data sovereignty and privacy regulations.
    
    \item \textbf{Immutable infrastructure} \\
    There is a strong preference in software engineering and data engineering for immutable data structures \cite{helland2015immutability}. 
    Immutable containers is a security best practice that simplifies container management and deployment in CI/CD pipelines \cite{tak2017understanding, rice2020container}, and
    Kubernetes works with immutable pods and deployments to make it easy for application developers to build auto-scaling and self-healing micro services.
    We would argue that the same immutability concept can and should be extended to lower levels of the infrastructure stack and, where possible, these infrastructure components should be "treated like cattle not pets''  \cite{morris2016infrastructure} \footnote{\url{http://cloudscaling.com/blog/cloud-computing/the-history-of-pets-vs-cattle/}}, i.e. 
    no ongoing maintenance through complicated and possibly customised patching and updating, just complete cluster replacement every time we need to make a change.
    Immutable infrastructure is fully knowable and directly documented through source code  \cite{mikkelsen2019immutable}.
    By eliminating configuration drift, immutable infrastructures make it easier for platform teams to maintain consistency between different environments and reduce security risks \cite{niculicea2019securing, kumara2021s}.

    \item \textbf{Automate whenever possible} \label{subsubsec:automate everything} \\
    To minimise configuration errors and maximise developer efficiency, we need to be exploit automation as much as possible. 
    We should seek to automate the entire stack, including cloud networking, storage, virtual machines, Kubernetes, and containerised data \& AI tools, ideally with a single command line tool. 
    Automation supported by human-in-the-loop admission controls before actual system deployment, for example a human developer explicitly accepting the risks found by vulnerability scanners,
    provides the best chance to avoid configuration errors that are so prevalent in cloud applications.

    \item \textbf{Low-friction enterprise ICT integration} \label{subsubsec:integrated} \\ 
    We aim to solve a whole problem for data and AI teams and provide a frictionless development experience.  
    This means \texttt{sskuba} clusters need to be well integrated with enterprise services such as single sign-on, hardened containers, public key infrastructure and logging. This integration breaks down barriers between different ICT teams, facilitating a seamless flow of data across operational and decision support systems and enabling the organisation to become more data-driven.

    \item \textbf{Secure by design} \label{subsubsec:secure by design} \\
    Secure systems allow developers to deliver at the speed of relevance by providing the tools and insights to mitigate risks in a controlled, informed way. 
    As the slogan goes, the software industry needs more secure products, not more security products. 
    By adopting secure by design practices, we make it possible for 
    \texttt{sskuba} to be trusted to operate effectively even in contested cyber environments, from the platform layer and all the way up the software stack \cite{hu2021artificial, kim2020impact, kathikar2023assessing, thuraisingham2022secure}.

\end{enumerate}

\subsection{Core Technologies}\label{subsec:core-infrastructure} 
The core technologies underlying \texttt{sskuba} are now described. 
They are used to create a software factory so developers are able to create repeatable infrastructure several times a day. 
These core technologies provide abstraction and portability to deploy \texttt{sskuba} clusters anywhere, and these abstractions are designed to provide the ability to adjust to expected shifts in underlying cloud and hardware technologies over time. 

\subsubsection{Container-specific Host Operating System}\label{subsec:host-os} 
The NIST defines a container-specific host operating system  \cite{NIST800190} as a minimalist OS explicitly designed to only run containers, with all other services and functionality disabled, and with read-only file systems and other hardening practices employed. 
They are sometimes also known as immutable operating systems \cite{bohm2023immutable}. 
The use of container-specific operating system on host machines for containers is a security best practice \cite{rice2020container}.

The virtual machines in a \texttt{sskuba} cluster are provisioned with Talos Linux, a container-specific operating system, with the following features \cite{avaznejad2022disk}:
\begin{itemize}
    \item \textbf{Minimal} Talos consists of only the necessary binaries and services for managing the OS and Kubernetes, such as containerd and kubelet. There is no package manager to expand the system functionality during runtime.
    \item \textbf{Immutable} Talos mounts the root file system as read-only using the compressed read-only file system SquashFS and the system does not change post deployment. 
    This immutability eliminates configuration drift and helps to prevent threat actors from tampering with the host even if they are able to somehow gain access to the host.
    
    \item \textbf{API driven} In Talos, APIs perform system management tasks such as diagnostics, upgrading Talos and Kubernetes, retrieving kernel logs, and listing network interfaces. These APIs are developed using Google Remote Procedure Call (gRPC) and secured with Mutual TLS (mTLS).
    
    \item \textbf{Universally deployable} Talos works on different platforms to ensure that the same operations are executed in the same way regardless of the underlying environment. It is deployable on single board computers (e.g. Raspberry Pi), bare metal, cloud (e.g. Google Cloud Platform (GCP) and Azure), virtualised (e.g. Hyper-V and VMware), and local platforms. It uses WireGuard \cite{donenfeld2017wireguard} to facilitate stretch clusters across hybrid cloud environments through KubeSpan, 
    a patented technology that securely and seamlessly establishes fully encrypted connections between all cluster members, even when they are operating on entirely different networks and behind firewalls.
    
    \item \textbf{Hardened} Talos is hardened with Kubernetes configurations as recommended by Center for Internet Security (CIS) benchmarks\footnote{\url{https://www.cisecurity.org/benchmark/kubernetes}}, including secure settings for the control plane and worker nodes. The OS is built based on the recommended configuration of the Kernel Self Protection Project\footnote{\url{https://kspp.github.io}}, a design of the Linux kernel to protect itself against potential flaws.
\end{itemize}

Talos being minimal and immutable reduces the attack surface and protects the system against some attack vectors because vulnerable services that are not required for Kubernetes are removed from the system.
Furthermore, Talos has no shell or SSH access, and API calls are secured with mTLS authentication, and essential certificates for encrypted communications are short-lived. 
A comparison of Talos with alternative container-optimised operating systems can be found in \S\ref{subsubsec:container OS}.

\subsubsection{Programming Language based Infrastructure as Code}

We are now in an era where we treat infrastructure not merely as scripts, but as actual software \cite{fitzgerald2015infrastructure}.
Within that general setup, Programming-Language-based Infrastructure as Code (PL-IaC) has been growing in popularity \cite{sokolowski2024reliable} and our preferred PL-IaC framework is Pulumi. 
It supports expressing infrastructure components such as storage, networking, virtual machines, and security in a consistent cloud object model. This model leverages standard software engineering constructs such as methods, classes, and packages. It packages industry-standard security best practices and policies into reusable libraries, thus supporting standardisation of multi-cloud deployments across an enterprise. This core abstraction provides repeatability and reliability to \texttt{sskuba} clusters. It also opens up the realm of infrastructure to software engineers, providing relief to organisations that have a shortage of skills and resources in platform engineering.

Second generation Infrastructure as Code (IaC) tools like Ansible and Terraform rely on the developers learning yet another Domain Specific Language (DSL),  leading to higher cognitive load and locks out many of the software-defined patterns and practices to infrastructure. 
Comparison and benchmarking of IaC tools can be found in \cite{kalliomaa2024choosing, karlsson2023comparison}.
A comparison of Terraform and Pulumi can be found in Appendix~\ref{app:pulumi terraform}.

\subsubsection{Shared Services}\label{subsec:shared-services}
In essentially all practical deployments, a \texttt{sskuba} cluster operates within a system of systems and need to integrate with enterprise ICT services and be given an authority to operate. 
Shared services in \texttt{sskuba} 
are lightweight software services used by multiple \texttt{sskuba} clusters. Whenever possible, they are built on top of secure and hardened enterprise cloud landing zones. They facilitate the cascading flow of information security controls, allowing users to inherit as many controls as possible.  

These shared services enhance the security posture, enable supply chain traceability and assurance, and accelerate the process of security accreditation for the analytics application deployed. Two key shared services needed are:
\begin{itemize}
    \item \textbf{Secure Remote Access}. Teleport\footnote{\url{https://github.com/gravitational/teleport}} is used to provide secure access to \texttt{sskuba} cluster at the Kubernetes layer through the issuance of short-lived X.509 certificates. Teleport integrates with existing identity providers, such as Azure Active Directory (Azure AD) or Github Enterprise. We further use Teleport for providing ingress authentication to secure web application deployed within \texttt{sskuba} clusters. Teleport lays the groundwork to meet FedRAMP level requirements for infrastructure access, including support for the Federal Information Processing Standard FIPS 140-2. This will enable \texttt{sskuba} clusters to attain, in the Australian context, Essential 8 maturity Level 3 for privileged access management.
    
    \item \textbf{Offline Containers and Artifacts}. Some \texttt{sskuba} cluster needs to operate under controlled outbound traffic (limited egress), intermittent, permanent or emergency disconnection from the network or internet (air-gapped mode). This shared service provides the ability to
    package a chunk of the internet (packages, libraries, containers, git repositories, helm charts, etc) and then securely deliver all the files needed to run the entire \texttt{sskuba} cluster in disconnected environments.

\end{itemize}

\subsection{How \texttt{sskuba-ctl} works?}\label{subsubsec:how sskuba-ctl works} 

\begin{figure}[!htbp]
    \centering
    \includegraphics[width=0.99\textwidth]{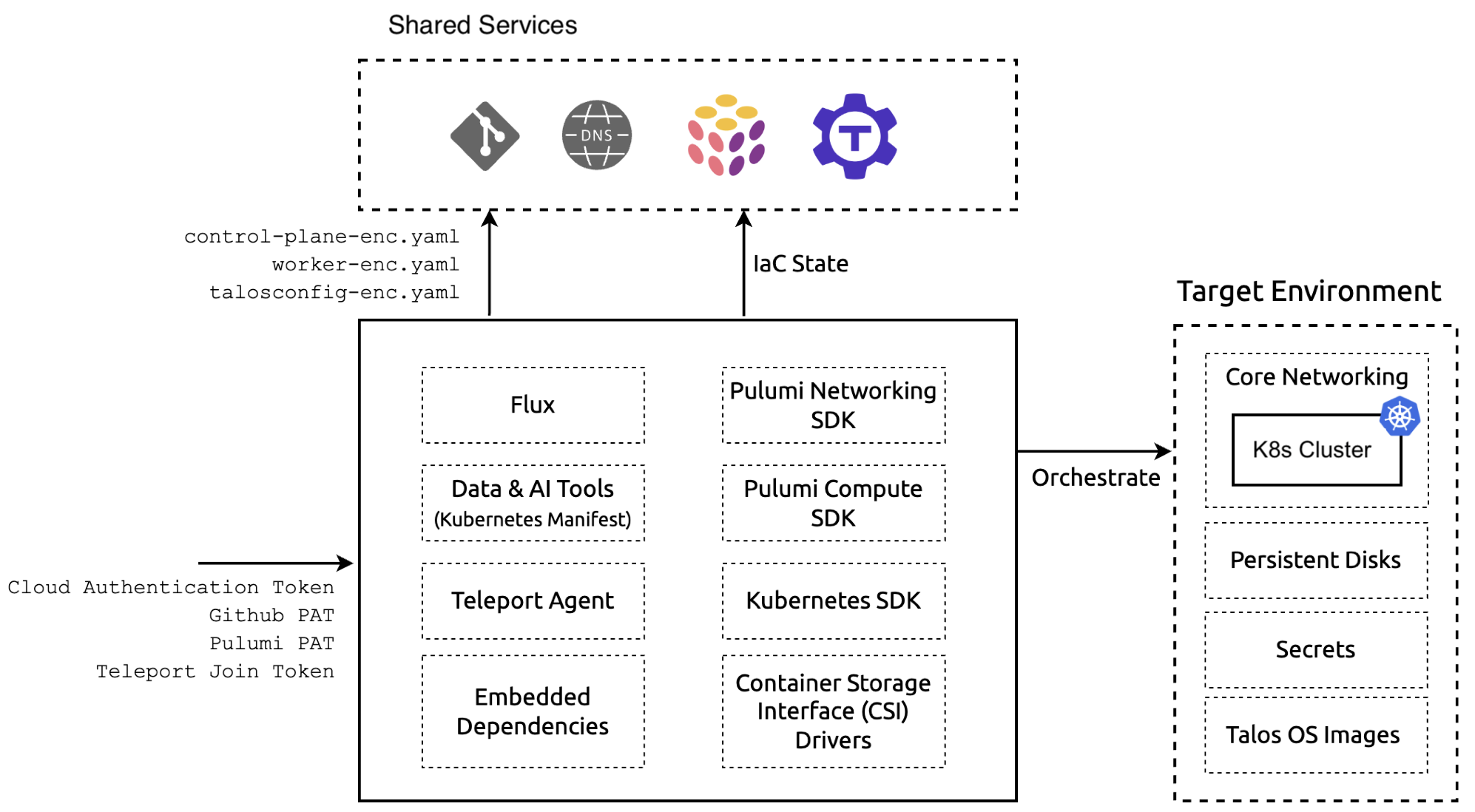}
    \caption{Logical Components of \texttt{sskuba-ctl}}
    \label{fig:sskuba-ctl-components}
\end{figure}

At the core of \texttt{sskuba} is a command line automation tool called \texttt{sskuba-ctl} written in Go, a general purpose programming language. 
It is a statically compiled binary available for multiple platforms (Linux, Mac and Windows), is configuration-driven and has been demonstrated to create a \texttt{sskuba} cluster in AWS, Azure and vSphere under ten minutes.
Figure \ref{fig:sskuba-ctl-components} shows the internal logical components of \texttt{sskuba-ctl}.

\subsubsection{Cluster Creation}
\begin{figure}
    \centering
    \includegraphics[width=0.8\textwidth]{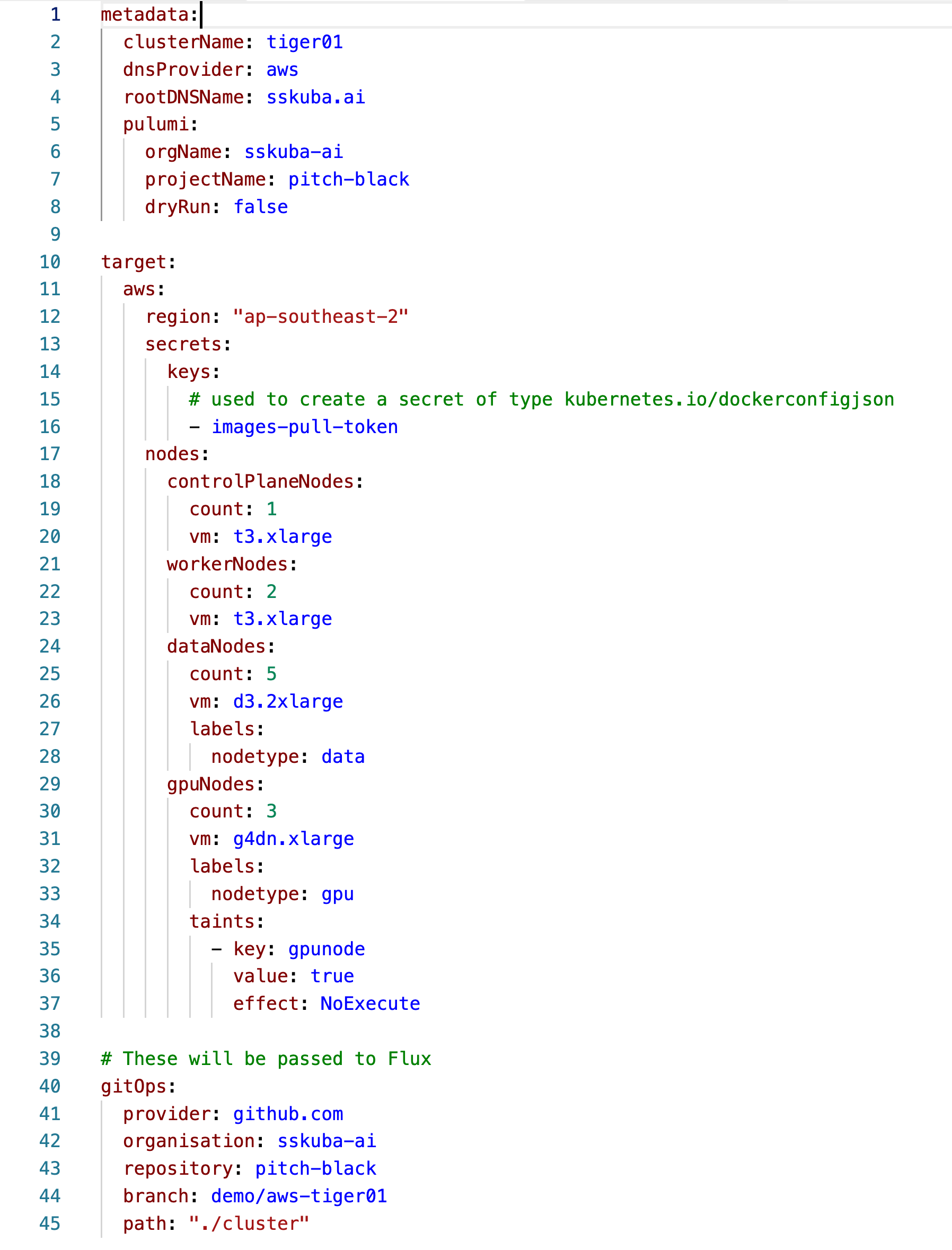}
    \caption{Sample configuration file - AWS Target}
    \label{fig:sample-configuration-file}
\end{figure}

\begin{figure}
    \centering
    \includegraphics[width=0.8\textwidth]{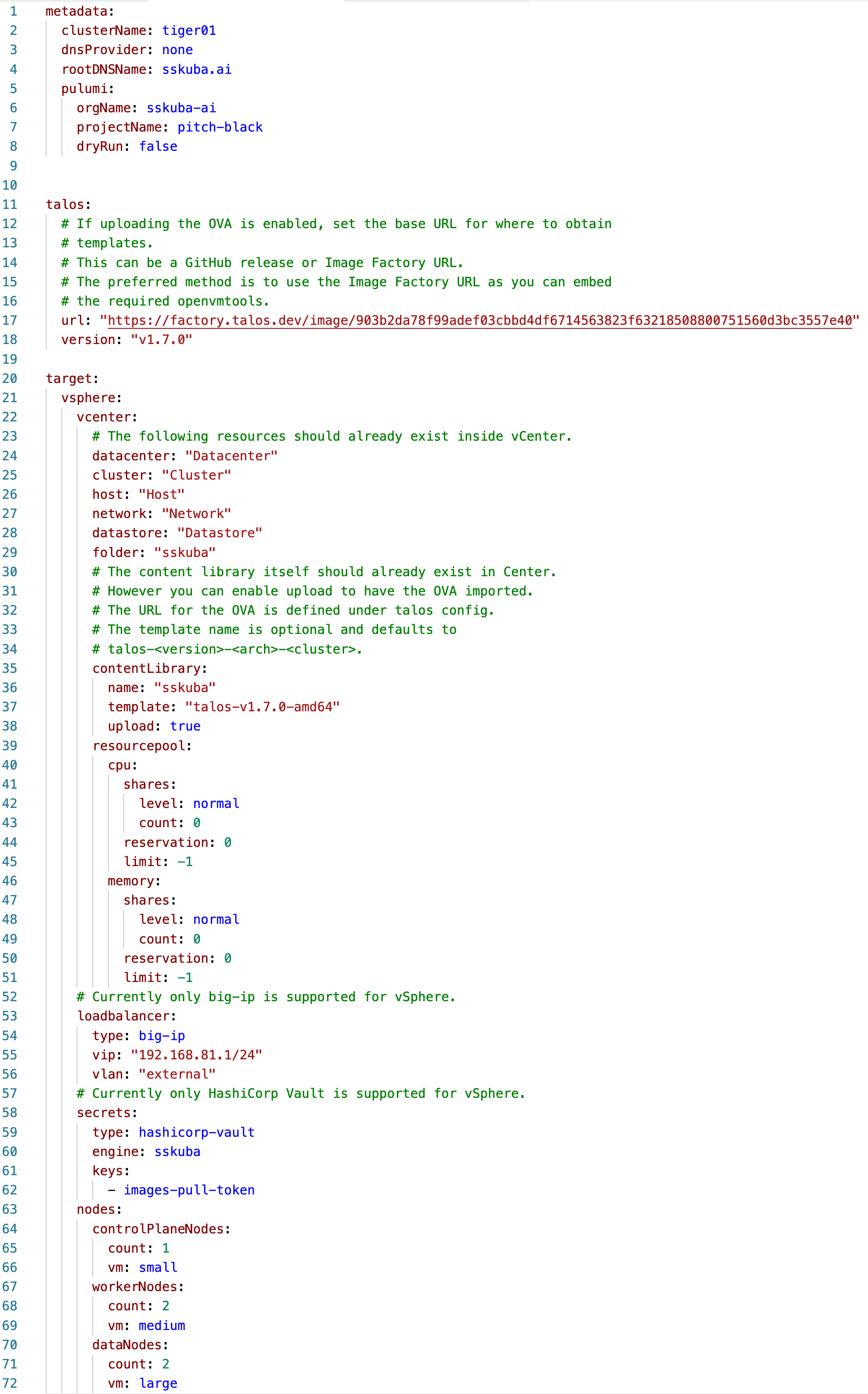}
    \caption{Sample configuration file - vSphere Target}
    \label{fig:sample-configuration-file-vSphere}
\end{figure}

The desired cluster setup is specified in a \texttt{sskuba} configuration file, which has three key sections. 
The \textit{Metadata} section provides the details for creating a unique fully qualified domain name in the target environment. 
The \textit{Target} section specifies the host environment (e.g. AWS, Azure, vSphere), including the number and type of virtual machines to be created. 
Finally, the \textit{GitOps} section provides a reference to a git repository that \texttt{sskuba} uses to bootstrap Flux, whereby \texttt{sskuba} installs the Flux controllers on the cluster and then configures the controllers to synch the cluster state from the specified git repository. 
Figures~\ref{fig:sample-configuration-file} and \ref{fig:sample-configuration-file-vSphere} show two example configuration files, one for AWS and one for vSphere. 

\begin{figure}[!htbp]
    \centering
    \includegraphics[width=0.9\textwidth]{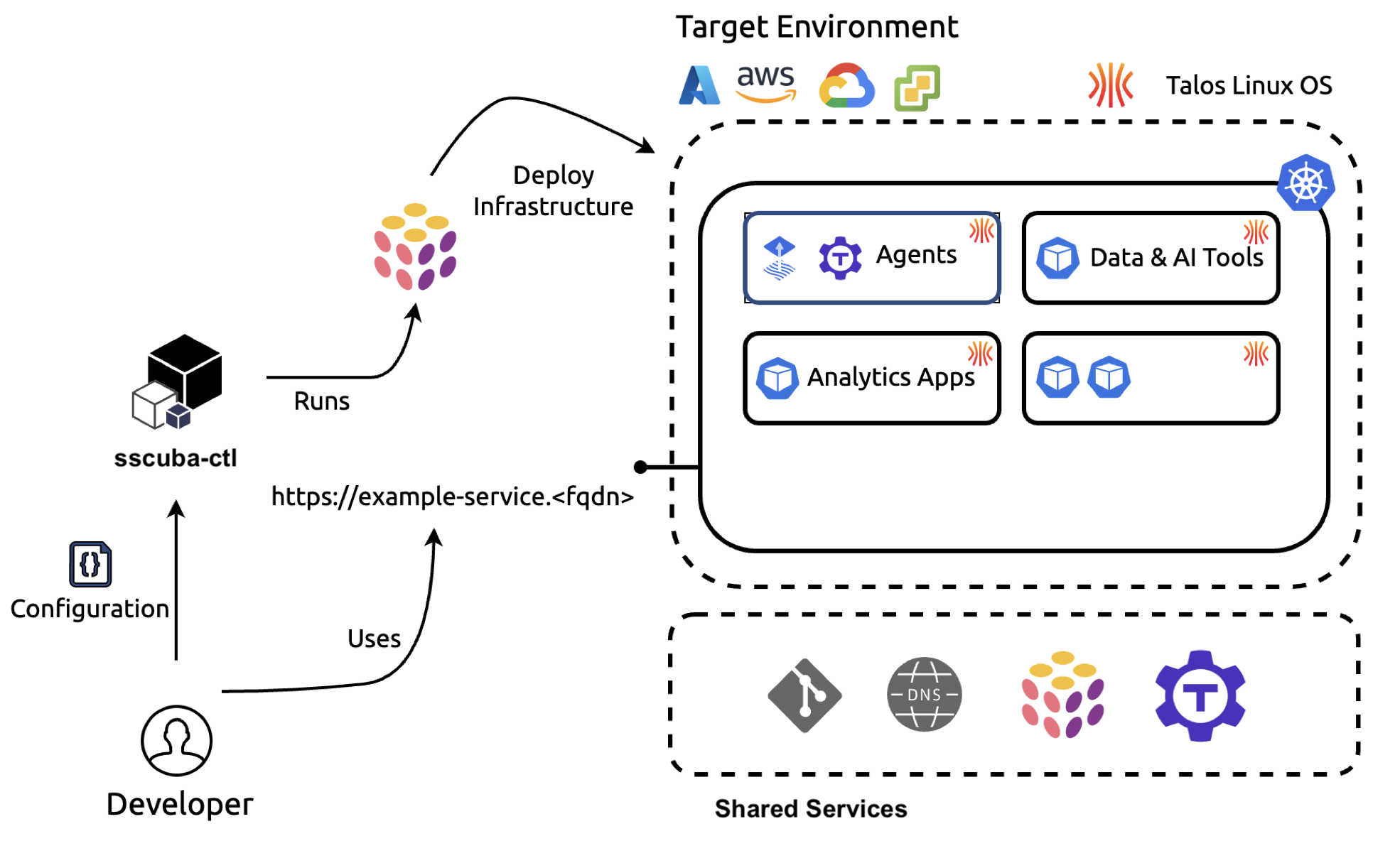}
    \caption{How does \texttt{sskuba-ctl} work?}
    \label{fig:sskuba-ctl-working}
\end{figure}

Figure~\ref{fig:sskuba-ctl-working} shows how \texttt{sskuba-ctl} works at a high level.
The cluster creation process can be run either locally on a developer machine or in a CI/CD runner with the command {\texttt sskuba-ctl apply -f cluster.yaml}. 
The exact steps taken in cluster creation are as follows:
\begin{itemize}
    \item \texttt{sskuba-ctl} starts by creating a Pulumi stack. Each \texttt{sskuba} cluster creates a new independent and isolated Pulumi stack. These stacks are organised into Pulumi projects and organisations for logical structure and management.
    
    \item The second step is creating the basic core networking required for a functioning Kubernetes cluster and the ability to route Talos OS Management traffic and ingress into the cluster. To achieve that, we create a Virtual Private Cloud (VPC), subnets, public IP, load balancer, security groups and network routes. The actual implementation details vary slightly depending on the target environment,  but the core capabilities remain the same. 
    
    \item The third step is creating virtual machines. The \texttt{sskuba} cluster is bootstrapped with new PKI certificates.
    These X.509 certificates, along with secrets, are used to generate Talos OS machine configuration files for the control plane and worker nodes. For GPU nodes, a modified worker node configuration is required to install and load NVIDIA GPU drivers. The respective configuration files are then used to create the virtual machines with the Talos OS images for the desired number of control plane, worker, and GPU nodes. After the machines have completed the boot process, the final step is to bootstrap etcd and apply labels and taints to the nodes and create a functioning Kubernetes cluster. 
    A \textit{kubeconfig} file is downloaded from the cluster for secure communication in break-glass scenarios. The machine configuration files contain certificate private keys and are encrypted using a new public/private key pair.
    
    \item The final step is to bootstrap FluxCD and deploy the data and AI tools with helm and kustomize. A teleport agent is deployed as a pod with the join token, which establishes connection and enable single sign-on to \texttt{sskuba} cluster with short-lived certificates. For friendly URLs a CNAME record is created in the hosted DNS with the value pointing to the public IP of the load balancer.
    Additional components for storage controllers, secrets and logging tools are also installed to enhance functionality.
    
\end{itemize}

Before \texttt{sskuba-ctl} can orchestrate and create the cluster, some configurations in shared services and target environment is required, which we cover next in \S\ref{subsubsec:shared services}. 

\subsubsection{Shared Services and Target Environments}\label{subsubsec:shared services}
The shared services listed in Figure~\ref{fig:sskuba-ctl-components} are now described in a bit more details.
\begin{itemize}
    \item \textbf{Domain Name Service (DNS)} - A hosted zone in the target environment (e.g. AWS Route53 or Azure DNS) needs to be set up to provide unique user-friendly URL for each \texttt{sskuba} cluster. 
    
    \item \textbf{Git repository} - A source code repository, which will be bootstrapped with FluxCD, that captures the desired state of core Kubernetes components, Data \& AI tools, and the analytics applications deployed inside the \texttt{sskuba} cluster. GitHub Personal Access Token (PAT) is necessary to read and write to a git repository configured for FluxCD.
    
    \item \textbf{Teleport} - 
    Teleport is an access management platform that acts as an identity-aware proxy that enables remote access to devices by generating short-lived certificates.
    A self-hosted Teleport server configured with an identity provider like Github or Azure Active Directory (AD) is needed to enable single sign-on in \texttt{sskuba} clusters. A Teleport join token is required by \texttt{sskuba-ctl} to automatically register a \texttt{sskuba} cluster on bootstrap. It can be generated by Teleport command line tool \textit{tctl} or by generating short-lived token in a CI/CD pipeline using Teleport MachineID.
    
    
\end{itemize}

The following are configurations or software components that are specific to target deployment environments, be they cloud or on-premises environments.
We give examples specific to AWS, Azure and vSphere to keep the discussion concrete. 

\begin{itemize} 

    \item \textbf{Cloud Authentication} - \texttt{sskuba-ctl} needs programmatic access to shared services and the target environment. For this, it uses native cloud tools (AWS CLI and Azure CLI) and individual PATs for local development. It uses OpenID Connect (OIDC) in CI/CD pipelines to orchestrate cloud resources providing short-lived dynamic tokens and eliminating the need for shared secrets and passwords.
    \item \textbf{Operating System Images} - Pre-configured Talos OS image templates are uploaded to Azure and AWS. A separate image with NVIDIA drivers is required for GPU nodes. 
    \item \textbf{Secrets} - Any Kubernetes secrets that are required at runtime are stored in native cloud secret managers or Hashicorp Vault for vSphere. An example is the token to authenticate container image registry. 
    \item \textbf{Storage} - Persistent volumes required for stateful workloads are dynamically created by deploying the cloud vendors' implementation of the Kubernetes Cloud Provider Interface with the CSI drivers. A default cloud specific storage-class is created for all \texttt{sskuba} clusters.
    \item \textbf{vSphere specific}: In the case of on-premises deployment Big-IP is used to provide load balancing; see Figure \ref{fig:sample-configuration-file-vSphere} for sample vSphere target configuration.

\end{itemize}

\begin{figure}[!htbp]
    \centering
    \includegraphics[width=1.02\textwidth]{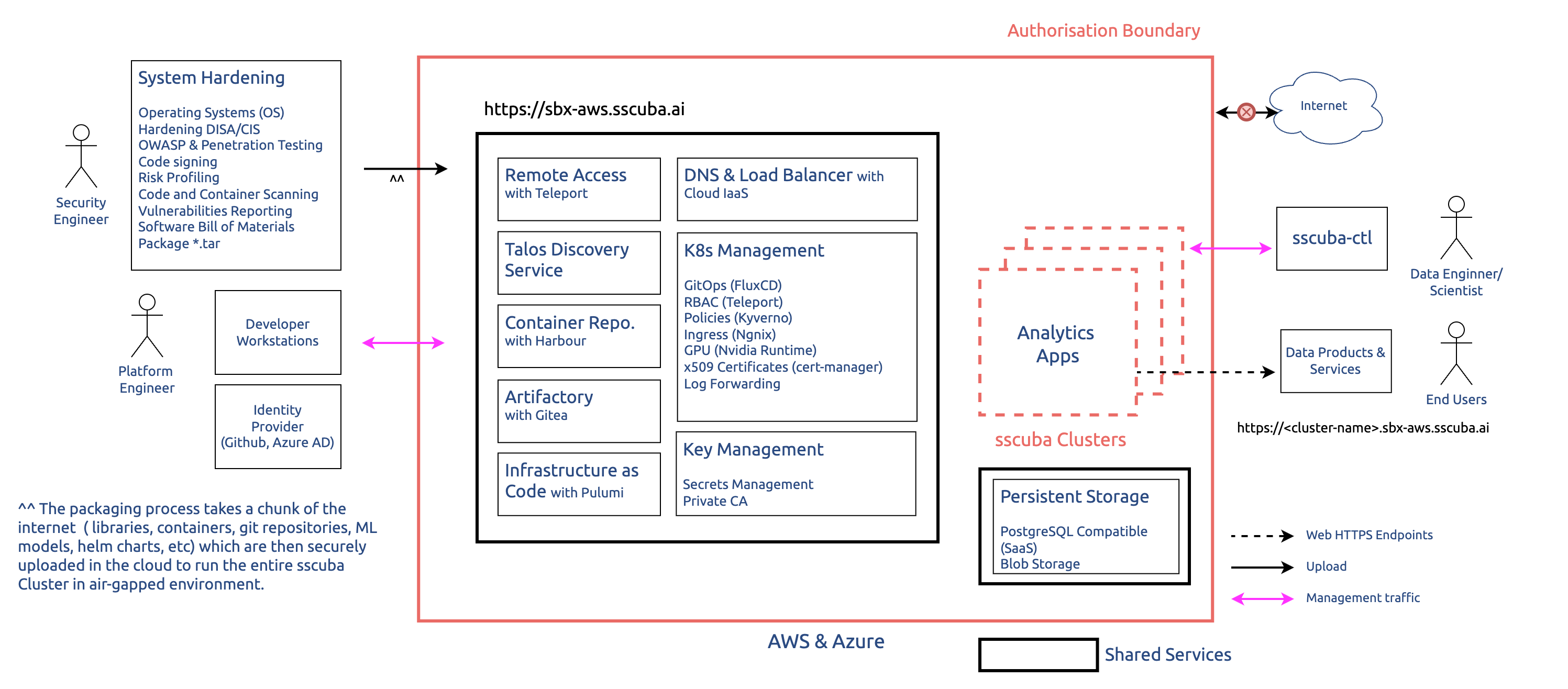}
    \caption{\texttt{sskuba} component technologies}
    \label{fig:sskuba technologies}
\end{figure}

\noindent Figure~\ref{fig:sskuba technologies} shows the component technologies in each of the shared services.

\subsection{Alternative Technologies}

In this section, we explore alternative choices to the set of core technologies selected for \texttt{sskuba} and explain the thinking process behind our decisions.

\subsubsection{Kubernetes Distributions and Managed Services}
The complexity that comes with setting up and managing Kubernetes clusters is the reason why many organisations use managed or prepackaged Kubernetes services from vendors, including major offerings like 
\begin{itemize}\itemsep1mm\parskip0mm
 \item Google Kubernetes Engine (GKE) 
 \item Amazon Elastic Kubernetes Service (EKS) 
 \item Azure Kubernetes Service (AKS)
 \item Red Hat OpenShift and its upstream / downstream platforms like Red Hat OKS and IBM CloudPak
 \item SUSE Rancher, which can be used to deploy the lightweight K3S as well as the RKE2 distribution designed for the US federal government sector.
\end{itemize}
While this is a reasonable position to take for low to medium-complexity organisations, it introduces risks like vendor lock-in and interoperability issues in large organisations with complex, distributed meshed ICT environments. 
Organisations that only require one stable and long-lived managed Kubernetes services will benefit from choosing one of the commerical offerings listed above.
For those that have a need for multiple Kubernetes clusters for different teams and/or different workloads, a lightweight solution is to write bespoke scripts using tools like kubeadm and clusterctl, perhaps together with Kubespray for declarative deployments, to get some consistency across the different Kubernetes clusters.
A more heavyweight solution is to work at a higher abstraction layer using something like Red Hat OKS or Red Hat OpenShift, which can provide a sufficiently common developer and user experience in hybrid cloud environments with offerings like Red Hat OpenShift Service on AWS, Azure Red Hat OpenShift, and Red Hat Device Edge (based on MicroShift, a lightweight Kubernetes distribution based on OpenShift).

Our proposed solution \texttt{sskuba-ctl} is inspired by \textit{eksctl} but is designed to be multi-cloud, lightweight and cost-effective to support large numbers of ephemeral AI workload in complex data mesh environments. Its key value-adds are
\begin{itemize}\itemsep1mm\parskip0mm
    \item integration with enterprise ICT services like identity and access management, public key infrastructure, and secrets management can be done consistently across deployment environments (cloud or on-premises);
    \item configurations that are "customer responsibilities" (from cloud vendors' perspective) will get secure and sensible defaults that are consistent across deployment environments (cloud or on-premises), which in turn will help with consistency in implementing and maintaining security controls \cite{shamim2020xi};
    \item a lightweight air-gapped Kubernetes cluster can be set up in denied, degraded, intermittent and limited (DDIL) environments.
\end{itemize}
Among the commercial offerings, \texttt{sskuba} is closest in design choices to SUSE Rancher, which provides two versions of Kubernetes for different deployment targets: K3S for single-node clusters and edge networks, and RKE2 for larger clusters and those that need strong security guarantees.
It is worth noting that RKE2 is a Kubernetes distribution specifically targeted for compliance with the US Government's Federal Information Processing Standard (FIPS), and it is DISA STIG-certified 
for environments with especially demanding security requirements.

\subsubsection{Container Orchestration and Management}
Kubernetes is, by now, the industry standard container orchestration and management solution.
Other alternatives exist, including Docker Swarm, Mesos, and various lightweight Kubernetes distributions.
Comparisons of these different alternatives can be found in \cite{malviya2022comparative, koziolek2023lightweight, telenyk2021comparison}.
For the intended applications of \texttt{sskuba}, we see no reason to adopt an alternative to Kubernetes.

\subsubsection{Container OS}\label{subsubsec:container OS}

In addition to Talos Linux, there are several (open-source) container operating systems available in the market.
\begin{itemize}\itemsep1mm\parskip0mm
 \item AWS Bottlerocket is a Linux-based operating system that is purpose-built by Amazon Web Services for running containers. It includes only the essential software required to run containers and is now generally available  as an Amazon Machine Image for Amazon Elastic Compute Cloud (EC2).\footnote{\url{https://aws.amazon.com/bottlerocket/}}
 
 \item Google Cloud Platform's Container-Optimized OS is an operating system image for Compute Engine VMs that is based on the open source Chromium OS project. Container-Optimized OS is the default node OS Image in Kubernetes Engine and other Kubernetes deployments on Google Cloud Platform.\footnote{\url{https://cloud.google.com/container-optimized-os}} 
 
 \item Red Hat Openshift runs on Red Hat CoreOS, which is based on Fedora CoreOS.
Azure does not have its own container-optimised OS like the other major cloud vendors but Azure supports running container OSs like Flatcar Container Linux\footnote{See \url{https://www.flatcar.org} and \url{https://fedoraproject.org/coreos/}}, which is also based on Fedora CoreOS.
\end{itemize}
From a cloud-agnostic and portability perspective, Flatcar Linux is really the only real alternative to Talos Linux. A comparison of Flatcar and Talos is provided in Appendix \ref{ap:talos vs flatcar}.
The summary is that while both Flatcar and Talos are designed for containerised workloads, Talos offers a more specialised, minimalistic, and API-driven approach specifically for Kubernetes environments whereas Flatcar provides a more traditional and versatile container-optimised Linux experience that can be used in a broader range of scenarios.
We have chosen Talos because we place a higher emphasis on security over versatility for \texttt{sskuba}'s intended use cases. 

\subsubsection{Serverless Computing}
Another approach to running secure, ephemeral AI workloads is to utilise serverless technologies, often referred to as Functions-as-a-Service (FaaS). Serverless computing is a contemporary execution model where users provide lightweight functions written in various programming languages. These functions are deployed to a platform that manages resource allocation, hosting, and execution automatically. AWS Lambda \cite{sbarski2017serverless} is an example of such a platform.

Serverless computing are built on lightweight isolation platforms acting as a middle ground between containers and full system virtualisation, by offloading functionality from the host kernel into an isolated guest environment. For example, Google’s gVisor can process many system calls in a user-mode \textit{Sentry} process, while AWS Firecracker runs a complete guest operating system within each microVM supported by a \textit{Jailer} process \cite{agache2020firecracker}. Both platforms boast rapid startup times, typically under 5 second and have a pay-per-second pricing model making them highly efficient and cost-effective solutions.

Serverless computing is particularly well-suited for certain AI workloads, such as serving machine learning models, including large language models (LLMs). Many commercial SaaS platforms, such as Fireworks AI\footnote{See \url{https://fireworks.ai}}, Baseten\footnote{See \url{https://www.baseten.co}} and RunPod\footnote{See \url{https://www.runpod.io}}, are already leveraging this approach to deliver fast model inference through developer-friendly APIs.

In terms of limitations, serverless platforms have specific resource constraints, such as memory limits (e.g., up to 10 GB for AWS Lambda), CPU allocation tied to memory, maximum execution time limits (e.g. 15 minutes for AWS Lambda), and restricted ephemeral storage (e.g. up to 10 GB on AWS Lambda with configuration). These functions typically run in isolated environments, often with restricted or limited network access. Communication with external services typically requires crossing the network boundary, introducing additional complexity. Due to their stateless nature and time limits, serverless functions are not ideal for long-running workloads like complex orchestrated data engineering processes and long-lived transaction-based database systems.

Finally, not all serverless platform provide the same level of security and isolation. 
For example, AWS ran serverless functions in Linux containers inside virtual machines, with each virtual machine dedicated to functions from a single tenant. In contrast, Azure will multiplex functions from multiple tenants on a single OS kernel in separate containers \cite{anjali2020blending}. As a result, a kernel bug could compromise inter-tenant security on Azure but not AWS.

\subsubsection{IDAM and PKI Integration Tools}
There are many solutions available for integrating a \texttt{sskuba} cluster with enterprise identity management and access (IDAM) systems and public key infrastructure (PKI) systems.
We picked Teleport for its flexibility and security.
A comparison of Teleport with alternatives like StrongDM, Hashicorp Boundary, Pomerium, and an argument for replacing SSH with Teleport in combination with Keycloak in modern software infrastructures, can be found in \cite{stocklin2022evaluating}.

\subsection{Limitations}
We end the description of \texttt{sskuba}'s self-service Kubernetes component by pointing out some of its limitations.
Architecture decisions made based on some of the design considerations listed in \S\ref{subsec:principles} have implications. 
Design principles like immutable infrastructure allow us to make simplifying assumptions about the system and its operating environment but these assumptions can also limit the potential use cases of \texttt{sskuba}. 
For example, the monolithic approach to \texttt{sskuba-ctl} with a baked-in container-native (immutable) host operating system provides a standardised, de-centralised and self-service data infrastructure to potentially hundreds of data teams across a large organisation but can introduce additional complexities when it comes to supporting bring-your-own-Kubernetes scenarios. Similarly GitOps based developer workflow provide auditability, continuous security monitoring and compliance at the cost of users who prefer low-code or no-code solutions. Finally, \texttt{sskuba} clusters are hyper optimised for stateless or short-lived data analytics use cases like executing data pipelines, training and serving ML models. So this limits running long lived distributed databases that need to be highly resilient, available or geo-distributed.  

\section{Data and AI Tool Kits}\label{sec:AI and data tools}

In this section, we describe the curated set of data and AI tool kits that come prepackaged in \texttt{sskuba} to help developers hit the ground running from day one.

\subsection{Design Considerations}\label{subsec:AI principles}
Before presenting the proposed solution in \S\ref{subsec:data-ai-tools}, we will start as usual by listing the design considerations.  

\subsubsection{Platform Design Principles}\label{subsec:tools principles} 
In addition to the principles listed in \S\ref{subsec:principles} for the self-service Kubernetes component, here are some additional principles and best practices when it comes to the AI and data tool kits component of \texttt{sskuba}. 

\begin{enumerate}
    \item \textbf{Open architecture} -- 
    AI and data analytics software stacks are increasingly converging onto the same set of features and functionalities. 
    We need to invest in an architectural blueprint with decoupled and modular software components based on functionality, modularity and communication patterns. The underlying components need to use open standards and protocols to ensure they are easy to integrate, upgrade and replace. This will result in open and stable architecture that is interoperable and adaptable to industry innovative without requiring wholesale changes.

    \item \textbf{External storage of system states} -- 
    Consistent with the immutability principle, in a stateless deployment model, the time needed to update the system is dramatically shorter than patching a stateful system.
    Therefore, it is best practice for data and system metadata to be clearly separated from tools and stored in persistent external storage and synchronised. 
    This design principle allows \texttt{sskuba} to be optimised for several data and AI use-cases like exploratory data analysis, serving of air-gapped Machine Learning models including LLMs, and execution of repeatable data pipelines at petabyte scale.

    \item \textbf{Balanced autonomy} -- 
    We need to provide a balance between the freedom to experiment and the need to deploy in secure standardised environments.
    We need to support rapid experimentation in development and testing, while transparently apply rigorous and centralised policies and standards as code moves closer to production. Simultaneously, we need to shift security to the left, giving developers early warnings and insights into vulnerabilities as early as possible. This balanced autonomy empowers data teams to quickly generate value through experimentation, while enhancing the organisation's cyber posture, enabling the creation of trusted data and AI products.

\end{enumerate}
   
\subsubsection{Sufficient Coverage of AI and ML Algorithms}

While numerous machine learning (ML) and AI algorithms are published annually, only a select few have significant impacts on real-world statistical practice.
How do we decide which algorithms to support in a modern AI and ML software stack?
One could start with the most popular and mature R and Python machine learning packages,\footnote{A popular curated list of machine learning frameworks, libraries and software is available at \url{https://github.com/josephmisiti/awesome-machine-learning}.}
but it would be useful to have a sense of what are considered core ML and AI algorithms and whether we have sufficiently good coverage over them, but within reason given unnecessary complexity can come with security risks.

The core set of ideas and algorithms as described in \cite{jordan2015machine} remains core and largely relevant today.
We can expand on those key concepts along two dimensions: (i) knowledge representation and reasoning formalisms, and (ii) machine learning principles and algorithms.
Figure~\ref{fig:KRR} in Appendix~\ref{app:krr} shows a map of mathematical structures that are useful for thinking about knowledge representation and reasoning (KRR) issues in AI and ML. 
It is built on top of a diagram in \cite{tegmark1998theory} and extended with our own understanding of historical and recent work across quite a few different fields of AI. 
Figure~\ref{fig:learning algorithms} in Appendix~\ref{app:ml algorithms} shows the major classes of algorithms in Machine Learning, organised around the associated induction principles and learning theory. 
From a design trade-off perspective, we would argue that the set of ML and AI algorithms supported in \texttt{sskuba} need to be 
\begin{itemize}\itemsep1mm\parskip0mm
 \item large enough to provide good, if not full, coverage of the topics covered in Figures~\ref{fig:KRR} and \ref{fig:learning algorithms}; and 
 \item small enough that cyber security controls and best practices can be effective in minimising attack surface and other security risks. 
\end{itemize}
The right balance is context-dependent, and we will need to constantly adjust to security and privacy issues like those considered in \S\ref{subsec:security and privacy}, and issues that come from the latest advances in AI like those considered in \S\ref{subsec:software 2.0}.

\subsubsection{Databases and Programming Languages}

When it comes to the design spectrum of database technologies, from best-of-breeds architecture \cite{StonebrakerMAHHH14} to monolithic enterprise data lakes \cite{hellerstein2012madlib} and all-in-one distributed storage engines that unifies data lakes and data warehouses \cite{lakshmanan2019google, levandoski2024biglake}, our preference is for best-of-breed database technologies because the `horses-for-courses' strategy fits best with our intended use case of supporting secure, ephemeral AI workloads in data mesh environments. This approach aligns with the RUM conjecture \cite{athanassoulis2016designing}, which posits that in the design of algorithms and data structures for organising and accessing data, one can optimise two out of the three factors -- read, update, and memory overhead -- at the expense of the third. While modern data system implementations can achieve partial optimisation across all three aspects, complete optimisation is infeasible due to the mutually exclusive nature of certain optimisation techniques.

A key question in a best-of-breed architecture is how do we choose the programming language(s) that will be used to `glue' and integrate the different components in the overall architecture into data-processing pipelines and workflows?

Every programming language and platform is essentially a derivative of one of three (equivalent) formulations of computability (see Figure~\ref{fig:KRR}).
Imperative languages like C and Java come from Turing machines and von Neumann machines.
Functional languages like Haskell and Lean with features like higher-order functions and interesting type systems, come from lambda calculus and higher-order logic.
Logic programming languages like Prolog, Answer Set Programming, and SQL that have built-in automated inference mechanisms come, in turn, from (fragments of) first-order logic and relational algebra. 
Most modern data-science languages sit at the intersection of two paradigms. 
For example, we have languages like R and Python with both imperative and functional features, and languages like PL\/pgSQL with both imperative and logic-programming features. From those languages evolved parallelised procedural SQL and Scala that run on parallel computers, for both scale-up and scale-out architectures.

Out of all these different choices, we prefer general-purpose programming languages that provide fundamental building blocks for creating functional, composable, secure and auditable software. 
Tools like Airflow and NiFi produce XML outputs that tightly couple business logic with technical boilerplate, which can make them hard to modularise and manage in source control. 
In contrast, tools like Dagster\footnote{\url{https://dagster.io}} and SQLGlot\footnote{\url{https://github.com/tobymao/sqlglot}} use general-purpose programming languages, thus creating more human readable code with short-feedback loop that enable build-low-deploy high processes, with development on local computers and deployment on (remote) distributed clusters.
A comparison of some key data and workflow orchestration tools is given in Appendix~\ref{app:data orchestration}.

\subsubsection{Data Security, Privacy and Confidential Computing}\label{subsec:security and privacy}

Cyber security and data security are closely related concepts that operate at different levels. 
Cyber security is primarily about controlling access to systems and data through different security protection mechanisms, from the physical network layer all the way to the application layer, and these security mechanisms come primarily in the form of encryption and digital signature algorithms, identity access management systems, and safe coding practices.
Data security is primarily about controlling access to data and, arguably more importantly, controlling what can and cannot be safely inferred from data that are provided to users.
While access control can be accomplished with cyber security and supporting functions like meta-data management, controlling what can be inferred from data, usually in the service of higher-level organisational goals like protecting user privacy and confidentiality, require a different class of technologies.

There are four confidential computing technologies that are applicable to a wide range of applications. 
Secure multiparty computation \cite{evans2018pragmatic}, through the use of secret
sharing schemes, permits multiple parties to jointly compute a function without divulging each party’s secret information to the others. 
Homomorphic encryption \cite{halevi2017homomorphic, lyubashevsky2024basic, li2022tutorial} tackles this issue from a different perspective, by encrypting sensitive data in a way that allows arithmetic operations to be performed directly on the encrypted data. 
Some privacy issues could arise through reverse-engineering the algorithm or query outcomes \cite{dinur2003revealing}. 
To address this, the differential privacy \cite{dwork2014algorithmic} framework can be used to provide individuals with guaranteed plausible deniability by adding suitably calibrated noise to query outcomes. 
Finally, federated learning is needed for multiple parties to jointly learn
a machine learning model from distributed data.
A detailed survey of these technologies and representative use cases in the Intelligence domain can be found in \cite{li2024privacy}.
We should aim to provide such confidential computing functionalities in the \texttt{sskuba} platform, either as standalone software libraries or through extensions to existing databases and languages like \cite{popa2011cryptdb}.

\subsubsection{Software 2.0 and LLMs}\label{subsec:software 2.0}

In his blog post\footnote{\url{https://karpathy.medium.com/software-2-0-a64152b37c35}} that introduced the term Software 2.0, Andrej Karpathy writes that unlike the `classical stack' of Software 1.0 where a programmer writes explicit instructions in languages like Python, C++, etc that are then compiled into a binary that performs useful work, Software 2.0 is written by computers in much more abstract, human unfriendly language, such as the weights of a neural network.
In particular, in Software 2.0, the human-supplied source code usually comprises (i) a high-level (mathematical) statement of what a good program looks like; (ii) a dataset of good and sometimes bad examples of the program's behavior, and (iii) a neural network architecture, usually with many layers and up to billions of parameters, that gives the rough skeleton of the code. 
The computer then proceeds to solve the given optimisation problem to find a good program, in the form of actual weights for the given neural network architecture, that exhibits the desired behaviour.
This strategy, somewhat surprisingly, has given rise to remarkable progress across a range of hard problems, from Go \cite{mnih2015human} to protein folding and self-driving cars.
Across many application areas, in Karpathy's words, we are now thus left "with a choice of using a 90\% accurate model we understand, or a 99\% accurate model we don't."
The \texttt{sskuba} platform needs to provide tools to support Software 2.0-style developer experience, informed by studies like \cite{dilhara2021understanding}.

In particular, in Software 2.0, data labelling and data engineering are the key to success and appropriate tools need to be provided for large-scale data wrangling across streaming and batch settings for structured, semi-structured and unstructured data. 
In addition, the platform should also support for data labelling as a first-class object in user interface designs \cite{ratner2019role, ratner2016data}.

With the maturation of code generators like GitHub Co-pilot and integrated programming assistants like Devin\footnote{\url{https://devin.ai}} and Replit\footnote{\url{https://replit.com}} based on large language models, we also need to consider how the \texttt{sskuba} platform can support and integrate with such tools in developer-friendly MLOps, LLMOps and AgentOps workflows \cite{tantithamthavorn2025mlops}.

Finally, programs generated using the Software 2.0 methodology can have unexpected failure modes, and there are now many widely reported canonical examples \cite{narayanan2024ai, christian2021alignment}.
Support for a comprehensive AI test and evaluation tool kits, which includes checklists \cite{kapoor2024reforms} and tools \cite{longpre2024responsible}, need to be provided in any modern data and AI platform.

\subsection{Data \& AI Platform Reference Implementation}\label{subsec:data-ai-tools}
\begin{figure}[H]
    \centering
    \includegraphics[width=0.6\textwidth]{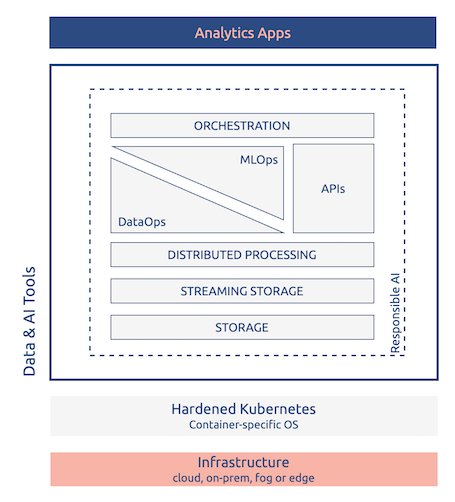}
    \caption{Data \& AI tool kits in \texttt{sskuba}}
    \label{fig:sskuba-capabilities}
\end{figure}

We are now ready to present a reference implementation of the data and AI tool kits that come preloaded in every \texttt{sskuba} cluster.
The tool kits is made up of a curated set of modern, cloud-native, fit-for-purpose software tools required to build, train, and deploy AI applications at scale. 
As per the common architecture practice, these tools can be logically grouped into seven capabilities: storage, distributed processing, DataOps, MLOps, streaming storage, orchestration, and APIs. These capabilities are illustrated in the Figure \ref{fig:sskuba-capabilities}.

Each capability comprises one or more component classes, which provide industry-standard features. For example, object storage, vector databases, and graph databases are all component classes within the storage capability.
Within each component class, we can select concrete implementations of the particular Data \& AI tool. For example, Neo4j and JanusGraph are components of the Graph component class within the storage capability.

\begin{figure} 
    \centering
    \includegraphics[width=1.05\textwidth]{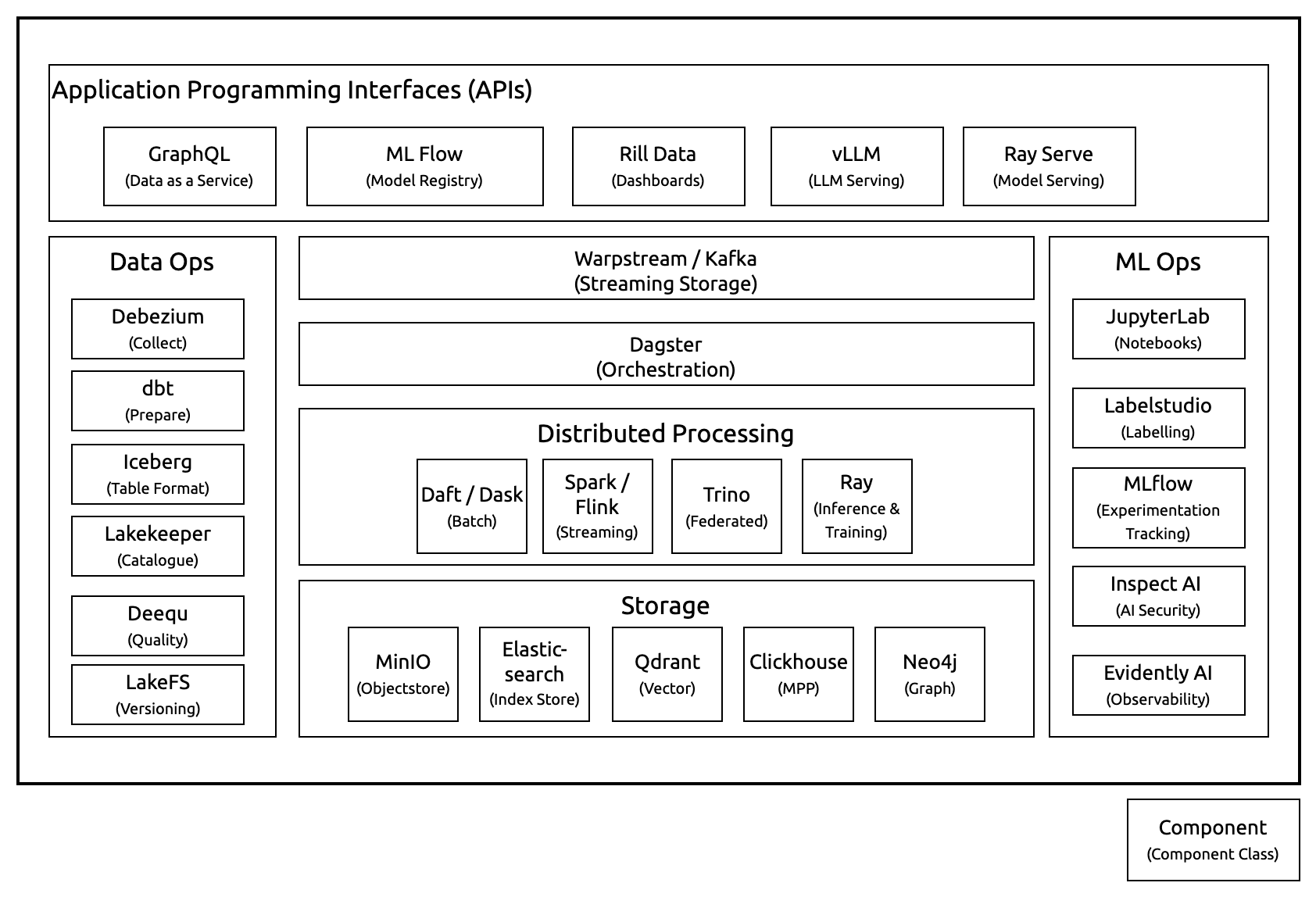}
    \caption{Reference implementation of \texttt{sskuba}'s data and AI tool kits}
    \label{fig:data-ai-tools-reference}
\end{figure}

Most organisations pick different software tools based on what works best for their teams and use cases -- it is a mix of art and science. 
Given the rapidly changing AI and data landscape as can be seen in Appendix~\ref{app:data AI landscape}, we will not do a detailed comparison of major technology choices in this paper. 
However, in the spirit of making things concrete, we show in Figure \ref{fig:data-ai-tools-reference} the selection of tools that we have used to test and build out the integrated architecture of \texttt{sskuba} over multiple use cases.
These technology choices reflect our understanding of the current state-of-the-art and are informed by the design considerations given in \S\ref{subsec:AI principles}.
But they are, so to speak, ``strong opinions loosely held'' -- while we have strong views on the superiority of these tools at the time of writing, our conviction is not so strong that we cannot change our minds when the facts and industry trends change.
We encourage the readers to do their own homework on these technology choices, which can now be done easily with modern large language models like ChatGPT and Perplexity.

\section{Discussion and Conclusion}

We have presented in this paper a data and AI platform called \texttt{sskuba} that seeks to overcome some of the key barriers to AI innovation and adoption in large enterprises.
The \texttt{sskuba} platform has two components:
\begin{itemize}\itemsep1mm\parskip0mm
    \item a self-service Kubernetes infrastructure that is designed to work within a data mesh architecture to enable secure and ephemeral AI workloads;
    \item a preloaded modern data and AI tool kits that developers can use to address a wide range of use cases at speed.
\end{itemize}

Our proposed self-service Kubernetes solution is designed to be multi-cloud, lightweight and cost-effective. 
Here are its key benefits:
\begin{itemize}\itemsep1mm\parskip0mm
    \item integration with enterprise ICT services like identity and access management and public key infrastructure can be done consistently across deployment environments (cloud or on-premises);
    
    \item configurations that are usually considered "customer responsibilities" (from cloud vendors' perspective) will get secure and sensible defaults that are consistent across deployment environments (cloud or on-premises);
    
    \item a lightweight air-gapped Kubernetes cluster can be easily set up for constrained ICT environments, including edge devices that operate in denied, degraded, intermittent and limited (DDIL) environments.
\end{itemize}

The availability of this secure and lightweight Kubernetes foundation makes it possible for us to build a best-of-breed data and AI tool kits that can evolve flexibly over time according to industry trends and best practices. That, we would argue, is what data and AI scientists and engineers ultimately need to deliver business benefits at the speed of relevance for their organisations.

\vspace{1em}
\addcontentsline{toc}{section}{References}
\bibliographystyle{plain}
\bibliography{references}

\begin{thebibliography}{10}

\bibitem{agache2020firecracker}
Alexandru Agache, Marc Brooker, Alexandra Iordache, Anthony Liguori, Rolf
  Neugebauer, Phil Piwonka, and Diana-Maria Popa.
\newblock Firecracker: Lightweight virtualization for serverless applications.
\newblock In {\em 17th USENIX Symposium on Networked Systems Design and
  Implementation}, pages 419--434, 2020.

\bibitem{anjali2020blending}
Anjali, Tyler Caraza-Harter, and Michael~M Swift.
\newblock Blending containers and virtual machines: a study of {F}irecracker
  and g{V}isor.
\newblock In {\em Proceedings of the 16th ACM SIGPLAN/SIGOPS International
  Conference on Virtual Execution Environments}, pages 101--113, 2020.

\bibitem{athanassoulis2016designing}
Manos Athanassoulis, Michael~S Kester, Lukas~M Maas, Radu Stoica, Stratos
  Idreos, Anastasia Ailamaki, and Mark Callaghan.
\newblock Designing access methods: The {RUM} conjecture.
\newblock In {\em EDBT}, volume 2016, pages 461--466, 2016.

\bibitem{avaznejad2022disk}
Parinaz Avaznejad.
\newblock Disk encryption on {T}alos operating system.
\newblock Master's thesis, Aalto University, 2022.

\bibitem{blei2017variational}
David~M Blei, Alp Kucukelbir, and Jon~D McAuliffe.
\newblock Variational inference: A review for statisticians.
\newblock {\em Journal of the American Statistical Association},
  112(518):859--877, 2017.

\bibitem{bode2024towards}
Jan Bode, Niklas K{\"u}hl, Dominik Kreuzberger, and Carsten Holtmann.
\newblock Towards avoiding the data mess: Industry insights from data mesh
  implementations.
\newblock {\em IEEE Access}, 2024.

\bibitem{bohm2023immutable}
Sebastian Bohm and Guido Wirtz.
\newblock Immutable operating systems: A survey.
\newblock In {\em CEUR Workshop Proceedings}, pages 52--60, 2023.

\bibitem{boyd2004convex}
Stephen Boyd and Lieven Vandenberghe.
\newblock {\em Convex Optimization}.
\newblock Cambridge University Press, 2004.

\bibitem{CSA2024}
Jon-Michael Brook, Randall Brooks, Alex Getsin, Vic Hargrave, Laura Kenner,
  Michael Morgenstern, Stephen Pieraldi, and Michael Roza.
\newblock
  \href{https://cloudsecurityalliance.org/artifacts/top-threats-to-cloud-computing-2024}{Top
  Threats to Cloud Computing 2024}.
\newblock Technical report, Cloud Security Alliance, 2024.

\bibitem{bryant2024kubernetes}
Lincoln Bryant, Robert~W Gardner, Fengping Hu, David Jordan, and Ryan~P Taylor.
\newblock Kubernetes deployment options for on-prem clusters.
\newblock {\em arXiv:2407.01620}, 2024.

\bibitem{burns15}
Brendan Burns.
\newblock How kubernetes changes operations.
\newblock {\em ;login:}, 40, 2015.

\bibitem{burns2016borg}
Brendan Burns, Brian Grant, David Oppenheimer, Eric Brewer, and John Wilkes.
\newblock Borg, {O}mega, and {K}ubernetes.
\newblock {\em Communications of the ACM}, 59(5):50--57, 2016.

\bibitem{cesa2006prediction}
Nicolo Cesa-Bianchi and G{\'a}bor Lugosi.
\newblock {\em Prediction, Learning, and Games}.
\newblock Cambridge University Press, 2006.

\bibitem{christian2021alignment}
Brian Christian.
\newblock {\em The Alignment Problem: How can Machines Learn Human Values?}
\newblock Atlantic Books, 2021.

\bibitem{coleman2016can}
Shirley Coleman, Rainer G{\"o}b, Giuseppe Manco, Antonio Pievatolo, Xavier
  Tort-Martorell, and Marco~Seabra Reis.
\newblock How can {SME}s benefit from big data? {C}hallenges and a path
  forward.
\newblock {\em Quality and Reliability Engineering International},
  32(6):2151--2164, 2016.

\bibitem{dehghani2022data}
Zhamak Dehghani.
\newblock {\em Data Mesh}.
\newblock Marcombo, 2022.

\bibitem{dietterich2000ensemble}
Thomas~G Dietterich.
\newblock Ensemble methods in machine learning.
\newblock In {\em Workshop on Multiple Classifier Systems}, pages 1--15.
  Springer, 2000.

\bibitem{dilhara2021understanding}
Malinda Dilhara et~al.
\newblock Understanding software-2.0: A study of machine learning library usage
  and evolution.
\newblock {\em ACM Transactions on Software Engineering and Methodology},
  30(4):1--42, 2021.

\bibitem{dinur2003revealing}
Irit Dinur and Kobbi Nissim.
\newblock Revealing information while preserving privacy.
\newblock In {\em Proceedings of the ACM Symposium on Principles of Database
  Systems}, pages 202--210, 2003.

\bibitem{donenfeld2017wireguard}
Jason~A Donenfeld.
\newblock Wireguard: Next generation kernel network tunnel.
\newblock In {\em NDSS}, pages 1--12, 2017.

\bibitem{donoho201750}
David Donoho.
\newblock 50 years of data science.
\newblock {\em Journal of Computational and Graphical Statistics},
  26(4):745--766, 2017.

\bibitem{dwork2014algorithmic}
Cynthia Dwork and Aaron Roth.
\newblock The algorithmic foundations of differential privacy.
\newblock {\em Foundations and Trends{\textregistered} in Theoretical Computer
  Science}, 9(3--4):211--407, 2014.

\bibitem{evans2018pragmatic}
David Evans, Vladimir Kolesnikov, and Mike Rosulek.
\newblock A pragmatic introduction to secure multi-party computation.
\newblock {\em Foundations and Trends{\textregistered} in Privacy and
  Security}, 2(2-3):70--246, 2018.

\bibitem{farmer2008seven}
William~M Farmer.
\newblock The seven virtues of simple type theory.
\newblock {\em Journal of Applied Logic}, 6(3):267--286, 2008.

\bibitem{ferry2018cloudmf}
Nicolas Ferry, Franck Chauvel, Hui Song, Alessandro Rossini, Maksym Lushpenko,
  and Arnor Solberg.
\newblock Cloud{MF}: {M}odel-driven management of multi-cloud applications.
\newblock {\em ACM Transactions on Internet Technology}, 18(2):1--24, 2018.

\bibitem{fitzgerald2015infrastructure}
Brian Fitzgerald, Nicole Forsgren, Klaas-Jan Stol, Jez Humble, and Brian Doody.
\newblock Infrastructure is software too!
\newblock {\em SSRN 2681904}, 2015.

\bibitem{freund1997decision}
Yoav Freund and Robert~E Schapire.
\newblock A decision-theoretic generalization of on-line learning and an
  application to boosting.
\newblock {\em Journal of Computer and System Sciences}, 55(1):119--139, 1997.

\bibitem{goedegebuure2024data}
Abel Goedegebuure, Indika Kumara, Stefan Driessen, Willem-Jan Van Den~Heuvel,
  Geert Monsieur, Damian~Andrew Tamburri, and Dario~Di Nucci.
\newblock Data mesh: a systematic gray literature review.
\newblock {\em ACM Computing Surveys}, 57(1):1--36, 2024.

\bibitem{vogels2006learning}
Jim Gray and Werner Vogels.
\newblock Learning from the {A}mazon technology platform.
\newblock {\em ACM Queue}, 4, 2006.

\bibitem{habbal2024artificial}
Adib Habbal, Mohamed~Khalif Ali, and Mustafa~Ali Abuzaraida.
\newblock Artificial {I}ntelligence trust, risk and security management ({AI}
  trism): Frameworks, applications, challenges and future research directions.
\newblock {\em Expert Systems with Applications}, 240:122442, 2024.

\bibitem{halevi2017homomorphic}
Shai Halevi.
\newblock Homomorphic encryption.
\newblock In {\em Tutorials on the Foundations of Cryptography: Dedicated to
  Oded Goldreich}, pages 219--276. Springer, 2017.

\bibitem{helland2015immutability}
Pat Helland.
\newblock Immutability changes everything.
\newblock {\em Communications of the ACM}, 59(1):64--70, 2015.

\bibitem{hellerstein2012madlib}
Joseph~M Hellerstein, Christoper R{\'e}, Florian Schoppmann, Daisy~Zhe Wang,
  Eugene Fratkin, Aleksander Gorajek, Kee~Siong Ng, Caleb Welton, Xixuan Feng,
  Kun Li, et~al.
\newblock The {MAD}lib analytics library or {MAD} skills, the {SQL}.
\newblock {\em Proceedings of the VLDB Endowment}, 5(12), 2012.

\bibitem{hong2019overview}
Jiangshui Hong, Thomas Dreibholz, Joseph~Adam Schenkel, and Jiaxi~Alessia Hu.
\newblock An overview of multi-cloud computing.
\newblock In {\em Workshop on Web, Artificial Intelligence and Network
  Applications}, pages 1055--1068. Springer, 2019.

\bibitem{hu2021artificial}
Yupeng Hu, Wenxin Kuang, Zheng Qin, Kenli Li, Jiliang Zhang, Yansong Gao,
  Wenjia Li, and Keqin Li.
\newblock Artificial intelligence security: Threats and countermeasures.
\newblock {\em ACM Computing Surveys}, 55(1):1--36, 2021.

\bibitem{Hurtgen2020}
Holger H\"{u}rtgen, Jan Kerkhof, and Manuel M\"{o}ller.
\newblock
  \href{https://www.mckinsey.com/capabilities/quantumblack/our-insights/rethinking-ai-talent-strategy-as-automated-machine-learning-comes-of-age}{Rethinking
  AI Talent Strategy as Automated Machine Learning Comes of Age}.
\newblock Technical report, McKinsey Analytics, 2020.

\bibitem{hutter2013probabilities}
Marcus Hutter, John~W Lloyd, Kee~Siong Ng, and William~TB Uther.
\newblock Probabilities on sentences in an expressive logic.
\newblock {\em Journal of Applied Logic}, 11(4):386--420, 2013.

\bibitem{jordan2015machine}
Michael~I Jordan and Tom~M Mitchell.
\newblock Machine learning: Trends, perspectives, and prospects.
\newblock {\em Science}, 349(6245):255--260, 2015.

\bibitem{kalliomaa2024choosing}
Niko Kalliomaa.
\newblock Choosing the right {I}a{C} tool for building reusable cloud
  infrastructure.
\newblock Master's thesis, University of Turku, 2024.

\bibitem{kapoor2024reforms}
Sayash Kapoor, Emily~M Cantrell, Kenny Peng, Thanh~Hien Pham, Christopher~A
  Bail, Odd~Erik Gundersen, Jake~M Hofman, Jessica Hullman, Michael~A Lones,
  Momin~M Malik, et~al.
\newblock {REFORMS}: Consensus-based recommendations for machine-learning-based
  science.
\newblock {\em Science Advances}, 10(18):eadk3452, 2024.

\bibitem{karlsson2023comparison}
Daniel Karlsson.
\newblock Comparison of infrastructure as code frameworks from a developer
  perspective.
\newblock Master's thesis, Linkoping University, 2023.

\bibitem{kathikar2023assessing}
Adhishree Kathikar, Aishwarya Nair, Ben Lazarine, Agrim Sachdeva, and Sagar
  Samtani.
\newblock Assessing the vulnerabilities of the open-source artificial
  intelligence landscape: A large-scale analysis of the {H}ugging {F}ace
  platform.
\newblock In {\em IEEE International Conference on Intelligence and Security
  Informatics}, pages 1--6, 2023.

\bibitem{kim2020impact}
Ashley~Hyowon Kim.
\newblock The impact of platform vulnerabilities in {AI} systems.
\newblock Master's thesis, Massachusetts Institute of Technology, 2020.

\bibitem{kim2017data}
Miryung Kim, Thomas Zimmermann, Robert DeLine, and Andrew Begel.
\newblock Data scientists in software teams: State of the art and challenges.
\newblock {\em IEEE Transactions on Software Engineering}, 44(11):1024--1038,
  2017.

\bibitem{kjorveziroski2022kubernetes}
Vojdan Kjorveziroski and Sonja Filiposka.
\newblock Kubernetes distributions for the edge: serverless performance
  evaluation.
\newblock {\em Journal of Supercomputing}, 78(11):13728--13755, 2022.

\bibitem{koller2009probabilistic}
Daphne Koller and Nir Friedman.
\newblock {\em Probabilistic Graphical Models: Principles and Techniques}.
\newblock The MIT Press, 2009.

\bibitem{koziolek2023lightweight}
Heiko Koziolek and Nafise Eskandani.
\newblock Lightweight kubernetes distributions: A performance comparison of
  microk8s, k3s, k0s, and microshift.
\newblock In {\em Proceedings of the 2023 ACM/SPEC International Conference on
  Performance Engineering}, pages 17--29, 2023.

\bibitem{kumara2021s}
Indika Kumara, Mart{\'\i}n Garriga, Angel~Urbano Romeu, Dario Di~Nucci, Fabio
  Palomba, Damian~Andrew Tamburri, and Willem-Jan van~den Heuvel.
\newblock The do’s and don’ts of infrastructure code: {A} systematic gray
  literature review.
\newblock {\em Information and Software Technology}, 137:106593, 2021.

\bibitem{lakshmanan2019google}
Valliappa Lakshmanan and Jordan Tigani.
\newblock {\em Google Big{Q}uery: The Definitive Guide}.
\newblock O'Reilly Media, 2019.

\bibitem{lattimore2020bandit}
Tor Lattimore and Csaba Szepesv{\'a}ri.
\newblock {\em Bandit algorithms}.
\newblock Cambridge University Press, 2020.

\bibitem{lecun2015deep}
Yann LeCun, Yoshua Bengio, and Geoffrey Hinton.
\newblock Deep learning.
\newblock {\em Nature}, 521(7553):436--444, 2015.

\bibitem{levandoski2024biglake}
Justin Levandoski et~al.
\newblock Big{L}ake: {B}ig{Q}uery's evolution toward a multi-cloud lakehouse.
\newblock In {\em Companion of the International Conference on Management of
  Data}, pages 334--346, 2024.

\bibitem{li2022tutorial}
Yang Li, Kee~Siong Ng, and Michael Purcell.
\newblock A tutorial introduction to lattice-based cryptography and homomorphic
  encryption.
\newblock {\em arXiv:2208.08125}, 2022.

\bibitem{li2024privacy}
Yang Li, Thilina Ranbaduge, and Kee~Siong Ng.
\newblock Privacy technologies for financial intelligence.
\newblock {\em arXiv:2408.09935}, 2024.

\bibitem{lloyd2007knowledge}
John~W. Lloyd.
\newblock Knowledge representation and reasoning in modal higher-order logic.
\newblock Technical report, Australian National University, 2007.

\bibitem{longpre2024responsible}
Shayne Longpre et~al.
\newblock The responsible foundation model development cheatsheet: A review of
  tools \& resources.
\newblock {\em arXiv:2406.16746}, 2024.

\bibitem{lyubashevsky2024basic}
Vadim Lyubashevsky.
\newblock Basic lattice cryptography: The concepts behind {K}yber ({ML}-{KEM})
  and {D}ilithium ({ML-DSA}).
\newblock {\em Cryptology ePrint Archive}, 2024.

\bibitem{malviya2022comparative}
Anshita Malviya and Rajendra~Kumar Dwivedi.
\newblock A comparative analysis of container orchestration tools in cloud
  computing.
\newblock In {\em International Conference on Computing for Sustainable Global
  Development}, pages 698--703, 2022.

\bibitem{meilua2024manifold}
Marina Meil{\u{a}} and Hanyu Zhang.
\newblock Manifold learning: What, how, and why.
\newblock {\em Annual Review of Statistics and Its Application}, 11, 2024.

\bibitem{mikkelsen2019immutable}
Anders Mikkelsen, Tor-Morten Gr{\o}nli, and Rick Kazman.
\newblock Immutable infrastructure calls for immutable architecture: Deploying
  a changeless architecture in the cloud.
\newblock In {\em Proceedings of the Hawaii International Conference on System
  Systems}, 2019.

\bibitem{mnih2015human}
Volodymyr Mnih et~al.
\newblock Human-level control through deep reinforcement learning.
\newblock {\em Nature}, 518(7540):529--533, 2015.

\bibitem{morris2016infrastructure}
Kief Morris.
\newblock {\em Infrastructure as code: {M}anaging servers in the cloud}.
\newblock O'Reilly, 2016.

\bibitem{mulchandani2022software}
Nand Mulchandani and John~NT Shanahan.
\newblock {\em Software-defined Warfare: Architecting the DOD's Transition to
  the Digital Age}.
\newblock Center for Strategic and International Studies, 2022.

\bibitem{muvsic2024digital}
Din Mu{\v{s}}i{\'c}, Jernej Hribar, and Carolina Fortuna.
\newblock Digital transformation with a lightweight on-premise {P}aa{S}.
\newblock {\em Future Generation Computer Systems}, 2024.

\bibitem{nambiar2022overview}
Athira Nambiar and Divyansh Mundra.
\newblock An overview of data warehouse and data lake in modern enterprise data
  management.
\newblock {\em Big Data and Cognitive Computing}, 6(4):132, 2022.

\bibitem{narayanan2024ai}
Arvind Narayanan and Sayash Kapoor.
\newblock {\em AI snake oil: What artificial intelligence can do, what it
  can’t, and how to tell the difference}.
\newblock Princeton University Press, 2024.

\bibitem{niculicea2019securing}
C-E. Niculicea.
\newblock Securing physical {IT} infrastructures through immutability.
\newblock Master's thesis, Lulea University of Technology, 2019.

\bibitem{nipkow2002isabelle}
Tobias Nipkow, Markus Wenzel, and Lawrence~C Paulson.
\newblock {\em Isabelle/HOL: a proof assistant for higher-order logic}.
\newblock Springer, 2002.

\bibitem{popa2011cryptdb}
Raluca~Ada Popa, Catherine~MS Redfield, Nickolai Zeldovich, and Hari
  Balakrishnan.
\newblock Crypt{DB}: Protecting confidentiality with encrypted query
  processing.
\newblock In {\em Proceedings of the 23rd ACM Symposium on Operating Systems
  Principles}, pages 85--100, 2011.

\bibitem{poulton2024kubernetes}
Nigel Poulton and Pushkar Joglekar.
\newblock {\em The {K}ubernetes Book}.
\newblock Independent, 2024.

\bibitem{ratner2016data}
Alexander~J Ratner, Christopher~M De~Sa, Sen Wu, Daniel Selsam, and Christopher
  R{\'e}.
\newblock Data programming: Creating large training sets, quickly.
\newblock {\em NeurIPS}, 29, 2016.

\bibitem{ratner2019role}
Alexander~J Ratner, Braden Hancock, and Christopher R{\'e}.
\newblock The role of massively multi-task and weak supervision in software
  2.0.
\newblock In {\em CIDR}, 2019.

\bibitem{rice2020container}
Liz Rice.
\newblock {\em Container Security: Fundamental Technology Concepts that Protect
  Containerized Applications}.
\newblock O'Reilly Media, 2020.

\bibitem{roy2023survey}
Padmaksha Roy, Jaganmohan Chandrasekaran, Erin Lanus, Laura Freeman, and Jeremy
  Werner.
\newblock A survey of data security: Practices from cybersecurity and
  challenges of machine learning.
\newblock {\em arXiv:2310.04513}, 2023.

\bibitem{sbarski2017serverless}
Peter Sbarski and Sam Kroonenburg.
\newblock {\em Serverless architectures on {AWS}: with examples using {AWS}
  {L}ambda}.
\newblock Simon and Schuster, 2017.

\bibitem{scholkopf2018learning}
Bernhard Scholkopf and Alexander~J Smola.
\newblock {\em Learning with kernels: Support vector machines, regularization,
  optimization, and beyond}.
\newblock MIT press, 2002.

\bibitem{Seth2024Navigating}
Dhruv Seth, Harshavardhan Nerella, Madhavi Najana, and Ayisha Tabbassum.
\newblock Navigating the {Multi}-{Cloud} {Maze}: Benefits, {Challenges}, and
  {Future} {Trends}.
\newblock {\em International Journal of Global Innovations and Solutions},
  2024.

\bibitem{shamim2020xi}
Md~Shazibul~Islam Shamim, Farzana~Ahamed Bhuiyan, and Akond Rahman.
\newblock {XI} commandments of {K}ubernetes security: A systematization of
  knowledge related to {K}ubernetes security practices.
\newblock {\em IEEE Secure Development}, pages 58--64, 2020.

\bibitem{skelton2019team}
M.~Skelton, M.~Pais, and R.~Malan.
\newblock {\em Team Topologies: Organizing Business and Technology Teams for
  Fast Flow}.
\newblock IT Revolution Press, 2019.

\bibitem{sokolowski2024reliable}
Daniel Sokolowski.
\newblock {\em Reliable Infrastructure as Code for Decentralized
  Organizations}.
\newblock PhD thesis, University of St. Gallen, 2024.

\bibitem{NIST800190}
Murugiah Souppaya, John Morello, and Karen Scarfone.
\newblock Application container security guide.
\newblock Technical Report SP 800-190, NIST, 2017.

\bibitem{stocklin2022evaluating}
Thomas~Thaulow St{\"o}cklin.
\newblock Evaluating ssh for modern deployments.
\newblock Technical report, Noroff University College, 2022.

\bibitem{StonebrakerMAHHH14}
Michael Stonebraker, Samuel Madden, Daniel~J. Abadi, Stavros Harizopoulos,
  Nabil Hachem, and Pat Helland.
\newblock The end of an architectural era: it's time for a complete rewrite.
\newblock In Michael~L. Brodie, editor, {\em Making Databases Work: the
  Pragmatic Wisdom of Michael Stonebraker}, pages 463--489. {ACM}, 2019.

\bibitem{sutton2018reinforcement}
Richard~S Sutton and Andrew~G Barto.
\newblock {\em Reinforcement Learning: An Introduction}.
\newblock MIT press, 2018.

\bibitem{tak2017understanding}
Byungchul Tak, Canturk Isci, Sastry Duri, Nilton Bila, Shripad Nadgowda, and
  James Doran.
\newblock Understanding security implications of using containers in the cloud.
\newblock In {\em USENIX ATC}, pages 313--319, 2017.

\bibitem{tantithamthavorn2025mlops}
Chakkrit~Kla Tantithamthavorn, Fabio Palomba, Foutse Khomh, and Joselito~Joey
  Chua.
\newblock {MLO}ps, {LLMO}ps, {FMO}ps, and beyond.
\newblock {\em IEEE Software}, 42(01):26--32, 2025.

\bibitem{tegmark1998theory}
Max Tegmark.
\newblock Is “the theory of everything” merely the ultimate ensemble
  theory?
\newblock {\em Annals of Physics}, 270(1):1--51, 1998.

\bibitem{telenyk2021comparison}
S.~Telenyk, Oleksii Sopov, Eduard Zharikov, and Grzegorz Nowakowski.
\newblock A comparison of {K}ubernetes and {K}ubernetes-compatible platforms.
\newblock In {\em International Conference on Intelligent Data Acquisition and
  Advanced Computing Systems}, volume~1, pages 313--317, 2021.

\bibitem{thuraisingham2022secure}
Bhavani Thuraisingham, Murat Kantarcioglu, and Latifur Khan.
\newblock {\em Secure Data Science: Integrating Cyber Security and Data
  Science}.
\newblock CRC Press, 2022.

\bibitem{van2014logic}
Johan Van~Benthem.
\newblock {\em Logic in Games}.
\newblock MIT press, 2014.

\bibitem{vaswani2017attention}
Ashish Vaswani, Noam Shazeer, Niki Parmar, Jakob Uszkoreit, Llion Jones,
  Aidan~N Gomez, {\L}ukasz Kaiser, and Illia Polosukhin.
\newblock Attention is all you need.
\newblock In {\em Advances in Neural Information Processing Systems}, pages
  5998--6008, 2017.

\bibitem{verma2015large}
Abhishek Verma, Luis Pedrosa, Madhukar Korupolu, David Oppenheimer, Eric Tune,
  and John Wilkes.
\newblock Large-scale cluster management at {G}oogle with {B}org.
\newblock In {\em European Conference on Computer Systems}, pages 1--17, 2015.

\bibitem{willems1995context}
Frans~MJ Willems, Yuri~M Shtarkov, and Tjalling~J Tjalkens.
\newblock The context-tree weighting method: Basic properties.
\newblock {\em IEEE Transactions on Information Theory}, 41(3):653--664, 1995.

\bibitem{wu2020comprehensive}
Zonghan Wu, Shirui Pan, Fengwen Chen, Guodong Long, Chengqi Zhang, and S~Yu
  Philip.
\newblock A comprehensive survey on graph neural networks.
\newblock {\em IEEE Transactions on Neural Networks and Learning Systems},
  32(1):4--24, 2020.

\end{thebibliography}

\newpage
\appendix
\section{Appendix}

\subsection{Certified Kubernetes Distributions and Installers}\label{app:k8s distributions}
Here are the CNCF certified Kubernetes offerings as of 12 Dec 2024 as listed on \url{https://www.cncf.io/training/certification/software-conformance/}.
\begin{figure}[H]
    \centering
    \includegraphics[width=1.0\textwidth]{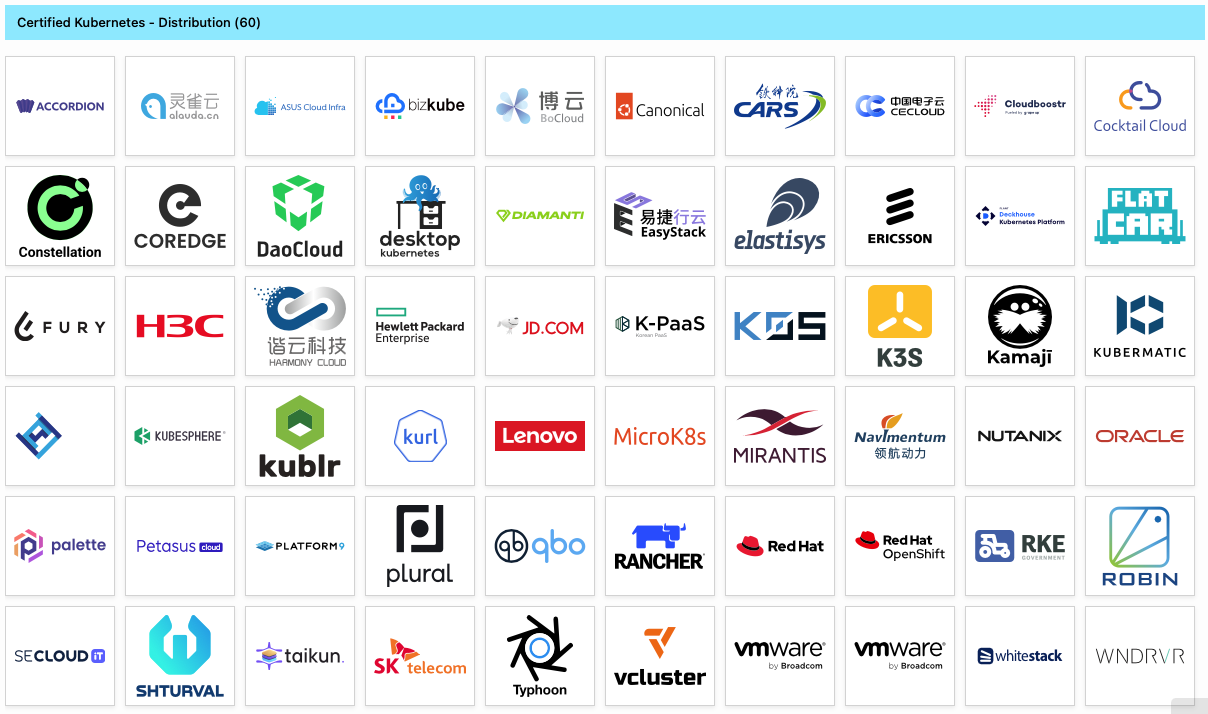}
    \includegraphics[width=1.0\textwidth]{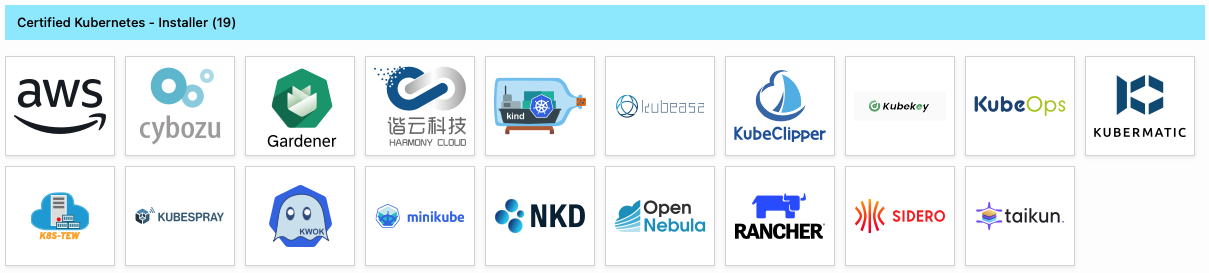}
        \includegraphics[width=1.0\textwidth]{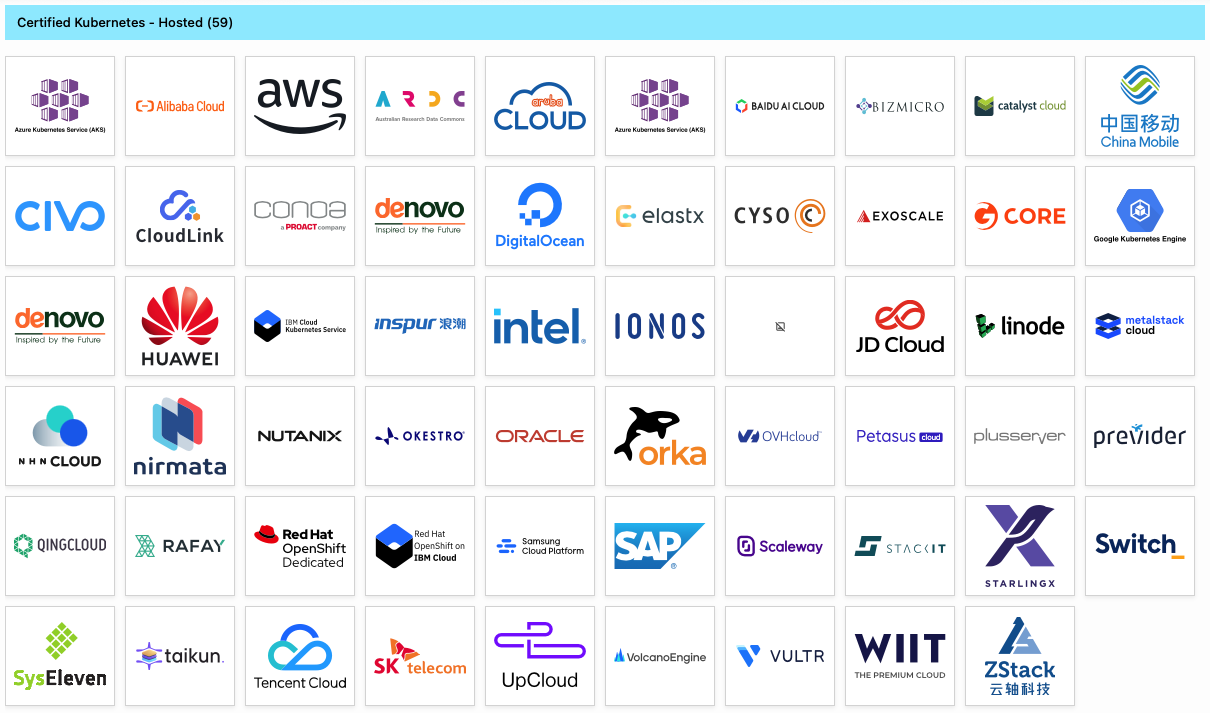}
\end{figure}

\newpage
\subsection{Data and AI Commercial Landscape}\label{app:data AI landscape}

\begin{figure}[H]
    \centering
    \includegraphics[width=0.8\textwidth]{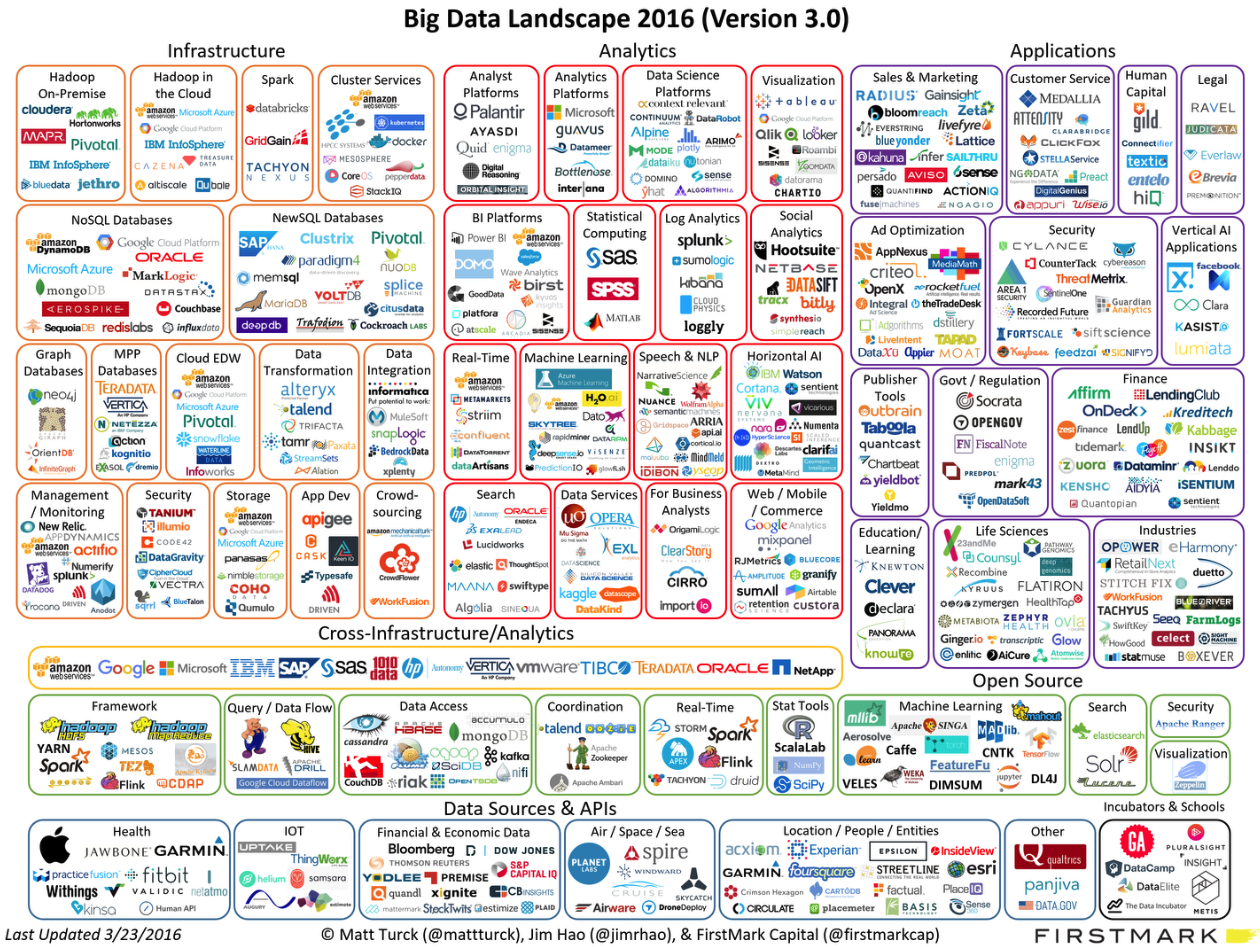} \\
    \caption{2016 Big Data Landscape; for details and commentaries, see \url{https://mattturck.com/big-data-landscape/}}
    \label{fig:mad landscape}
\end{figure}

\begin{figure}[H]
    \centering
    \includegraphics[width=1\textwidth]{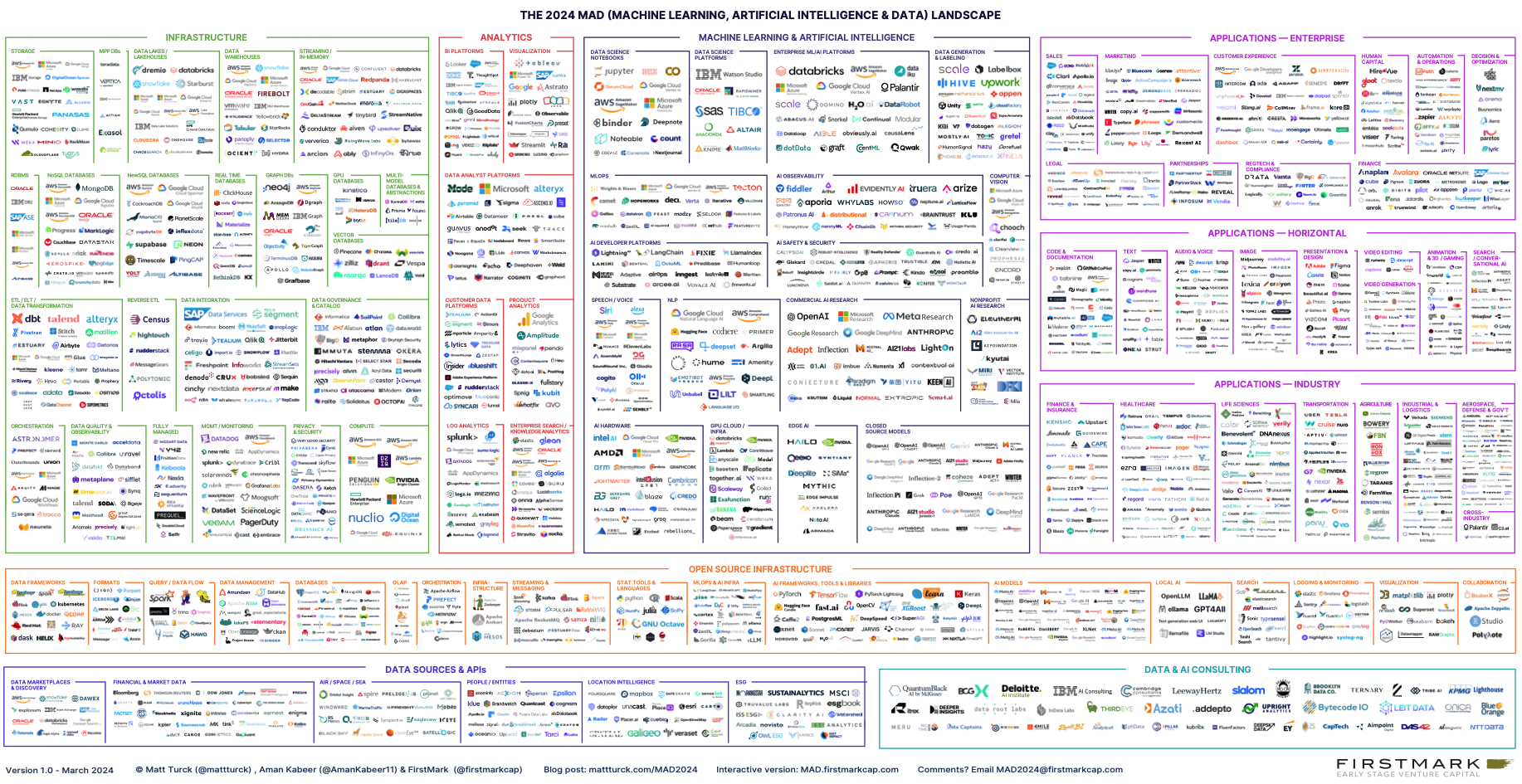}
    \caption{2024 Machine Learning, AI and Data Landscape; for details and commentaries on the latest trends, see  \url{https://mattturck.com/mad2024/}}
    \label{fig:mad landscape 2024}
\end{figure}

The number and variety of data and AI tools have exploded in the last 10-15 years.
Matt Turck of FirstMark Capital (\url{http://mattturck.com}) publishes an annual survey of the data and AI landscape.
Figure~\ref{fig:mad landscape} shows the landscape as of 2016, and Figure~\ref{fig:mad landscape 2024} shows the landscape for 2024.

\newpage
\subsection{Comparison of Pulumi and Terraform}\label{app:pulumi terraform}
The following was generated by \url{Perplexity.ai}, modulo some minor editing. The text and its sources have been checked by the authors. 
Pulumi and Terraform are two prominent tools in the Infrastructure as Code landscape, each catering to different developer preferences and use cases. Below is a comparison based on several key factors.
\begin{enumerate}\itemsep0mm\parskip0mm
 \item Configuration Language
 \begin{itemize}
     \item Pulumi: Supports multiple programming languages including Python, JavaScript, TypeScript, Go, C\#, and YAML. This flexibility allows developers to use familiar languages, making it easier to integrate infrastructure code with application code.
     \item Terraform: Utilizes HashiCorp Configuration Language (HCL), which is a domain-specific language designed specifically for defining infrastructure. While HCL is powerful for its purpose, it requires learning a new syntax for those not already familiar with it.
  \end{itemize}
\item State Management
  \begin{itemize}
      \item Pulumi: Manages state through its own service by default, which can also be configured to use other cloud services. This allows for easier sharing of state across teams and includes features like policy enforcement.
     \item Terraform: Uses local JSON state files (terraform.tfstate) for managing state, requiring some manual handling. 
     It provides options for remote state management but generally requires more setup compared to Pulumi.
  \end{itemize}

\item Community and Ecosystem
  \begin{itemize}
    \item Pulumi: While newer, it has been rapidly growing and provides support for over 60 major cloud services. Pulumi can also utilize Terraform providers, expanding its ecosystem significantly.
    \item Terraform: Has a more established community with a vast library of modules and plugins created by users. Its maturity means there is a wealth of resources, tutorials, and community support available.
  \end{itemize}
  
\item Extensibility
  \begin{itemize}
      \item Pulumi: Offers Dynamic Provider Support, allowing it to create new providers and support features faster than Terraform. This is particularly beneficial for developers needing cutting-edge features quickly.
      \item Terraform: While extensible through community plugins, it may lag in supporting the latest cloud features compared to Pulumi. 
 \end{itemize}
 
\item Testing and Development Workflow
  \begin{itemize}
      \item Pulumi: Being able to use general-purpose programming languages allows developers to leverage existing testing frameworks for unit and integration testing, enhancing the reliability of their infrastructure code.
      \item Terraform: Primarily supports integration tests; unit testing is less straightforward due to the nature of HCL being a configuration language rather than a full programming language.
  \end{itemize}
  
\item Licensing
  \begin{itemize}
      \item Pulumi: Uses the Apache License 2.0, which is permissive and straightforward for open-source contributions.
      \item Terraform: Transitioned to a Business Source License (BSL) in August 2023, which may impose restrictions on competitive use unless an agreement is signed with HashiCorp.
  \end{itemize}

\end{enumerate}

Choose Pulumi if:
\begin{itemize}\itemsep1mm\parskip0mm
    \item You prefer using familiar programming languages.
    \item You require dynamic provider support or rapid feature updates.
    \item Your team values integrated testing within the development workflow.
\end{itemize}
\noindent Choose Terraform if:
\begin{itemize}\itemsep1mm\parskip0mm
    \item You need a well-established tool with extensive community resources.
    \item Your infrastructure requirements are primarily based on traditional VMs or complex multi-cloud environments.
    \item You prefer a declarative approach that focuses solely on infrastructure without additional programming overhead.
\end{itemize}

In summary, both Pulumi and Terraform offer robust solutions for managing infrastructure as code but cater to different needs and preferences. The choice between them often depends on the specific requirements of your projects and your team's familiarity with programming languages versus configuration languages.

\newpage
\subsection{Comparison of FlatCar Linux and Talos Linux}\label{ap:talos vs flatcar}
The following was generated by \url{Perplexity.ai}, modulo some minor editing. The text and its sources have been checked by the authors. 
Flatcar Linux and Talos Linux are both container-optimized operating systems designed for running containerized workloads, but they have several key differences:
\begin{enumerate}\itemsep0mm\parskip1mm
\item Architecture and Design
 \begin{itemize} 
  \item Flatcar Linux is based on CoreOS, which in turn is derived from Gentoo Linux. It provides a more traditional Linux environment optimized for containers. \item Talos Linux, on the other hand, is a highly minimalistic and purpose-built OS designed specifically for running Kubernetes.
\end{itemize}

\item System Footprint -- Talos Linux has a significantly smaller footprint. 
 \begin{itemize}
     \item Talos has only 12 unique binaries in the system PATH.
    \item Flatcar has over 2300 binaries in the system PATH.
 \end{itemize}
This reduced footprint in Talos minimizes the attack surface and simplifies system maintenance.

\item Immutability and Security -- Both systems emphasize immutability, but Talos takes this concept further:
\begin{itemize}
   \item Talos has a more extensively read-only filesystem, with only specific files mounted into configuration paths. Talos 
   also implements features like secure boot and a hardened kernel configuration.
   
    \item Flatcar's /usr path is read-only, but many parts of the filesystem, including /, are not.
 \end{itemize}

\item Configuration and Management
 \begin{itemize} 
  \item Flatcar uses a declarative configuration approach with tools like Ignition and cloud-init. 
  \item Talos employs a gRPC API for all system configuration and management, removing the need for SSH access. This API-driven approach 
  simplifies automation and management, especially in bare metal environments.
 \end{itemize}

\item Update Mechanism -- Both systems use image-based updates:
 \begin{itemize}
  \item Flatcar uses an A/B partition scheme inspired by ChromeOS for updates and rollbacks.
  \item Talos uses a custom update mechanism that is claimed to provide more granular control over updates.
 \end{itemize}
 
\item Use Case Focus
 \begin{itemize} 
  \item Flatcar is more versatile and can be used for general container workloads, including but not limited to Kubernetes. 
  \item Talos is optimised for Kubernetes environments, integrating tightly with Kubernetes and not intended as a general-purpose Linux distribution.
 \end{itemize}
 
\item Community and Support -- Both projects emphasize open-source development and community engagement.
 \begin{itemize} 
  \item Flatcar has a larger community, being a direct successor to CoreOS. 
  \item Talos, being newer, has a smaller but growing community. 
  \end{itemize}
  
\end{enumerate} 

In summary, while both Flatcar and Talos are designed for containerized workloads, Talos offers a more specialized, minimalistic, and API-driven approach specifically for Kubernetes environments. Flatcar provides a more traditional and versatile container-optimized Linux experience that can be used in a broader range of scenarios.

\newpage
\subsection{Comparison of Dagster, Airflow and Nifi}\label{app:data orchestration}
The following was generated by \url{Perplexity.ai}, modulo some minor editing. The text and its sources have been checked by the authors. 
Dagster, Apache Airflow, and Apache NiFi are all popular tools in the data engineering ecosystem, but they serve different purposes and have distinct strengths. Let's compare these three tools based on their key features, use cases, and characteristics.

\paragraph{Dagster}
Dagster is a modern data orchestration platform that focuses on the entire data lifecycle, from development to production.
It is primarily designed for Python users, leveraging Python's popularity in the data engineering and analytics space.
Here are its key features:
\begin{itemize}\itemsep1mm\parskip0mm
 \item Asset-oriented approach, focusing on data assets rather than just tasks
 \item Strong support for local development, testing, and debugging
 \item Python-based Domain Specific Language (DSL) for defining workflows
 \item Built-in data validation and error handling
 \item Integration with ML frameworks like TensorFlow and PyTorch
 \item Emphasis on testing and reproducibility
\end{itemize}

\noindent Best for:
\begin{itemize}\itemsep1mm\parskip0mm
 \item Teams and projects looking for a comprehensive data engineering lifecycle management solution
 \item Teams and projects requiring strong data validation and error handling, and fine-grained control over data lineage
 \item Workflows involving machine learning pipelines
\end{itemize}

\paragraph{Apache Airflow} 
Airflow is a widely-used, task-based workflow orchestration platform.
It is primarily Python-based but offers more flexibility in terms of language support, allowing developers to integrate tasks written in different languages.
Here are its key features:
\begin{itemize}\itemsep1mm\parskip0mm
 \item Task-based workflow definition using Directed Acyclic Graphs (DAGs)
 \item Dynamic task generation
 \item Extensive library of built-in operators for common tasks
 \item Large community and ecosystem of plugins and integrations
 \item Web-based user interface for monitoring and managing workflows
\end{itemize}

\noindent Best for:
\begin{itemize}\itemsep1mm\parskip0mm
 \item Scheduling and monitoring complex batch workflows
 \item Teams with existing investments in the Apache ecosystem
 \item Projects requiring integration with a wide variety of tools and services
\end{itemize}

\paragraph{Apache NiFi} 
NiFi is a dataflow management and automation tool designed for real-time data ingestion and processing.
It takes a more language-agnostic approach compared to Dagster and Airflow.
Here are its key features:
\begin{itemize}\itemsep1mm\parskip0mm
 \item Visual interface for designing, controlling, and monitoring dataflows
 \item Real-time data flow capabilities
 \item Data provenance tracking
 \item Over 100 built-in processors for various data operations, and supports multiple languages for custom processors, including Java, Python, and others
 \item Supports a wide range of data formats and protocols
\end{itemize}

\noindent Best For:
\begin{itemize}\itemsep1mm\parskip0mm
 \item Real-time data ingestion and processing scenarios
 \item Projects requiring continuous data flow automation
 \item Teams needing visual dataflow design and monitoring
\end{itemize}

\paragraph{Comparison Table}
Here is a summary comparison.
\begin{table}[!htbp]
\begin{small}
\begin{tabular}{|c|p{3.3cm}|p{3.2cm}|p{3.2cm}|}
\hline
\textbf{Feature}	& \textbf{Dagster}	& \textbf{Apache Airflow}	& \textbf{Apache NiFi} \\
\hline \hline
Primary Focus	& Data assets and lifecycle management	& Task-based workflow orchestration	& Real-time dataflow automation \\
\hline
Architecture	& Pipeline-based	& Task-based	& Dataflow-based \\
\hline
Primary Language & Python & Python & Language-agnostic \\
\hline
Local Development	& Strong support	& Limited support	& Limited support \\
\hline
Data Validation	& Built-in	& Limited	& Limited \\
\hline
ML Integration	& Native support	& Via plugins	& Limited \\
\hline
Visual Interface	& Yes	& Yes	& Yes (extensive) \\
\hline
Real-time Processing	& Limited	& Limited	& Strong support \\
\hline
Community Size	& Growing	& Large	& Moderate \\
\hline
\end{tabular}
\end{small}
\end{table}

\paragraph{When to Choose Each Tool}
\begin{itemize}\itemsep1mm\parskip0mm
 \item Choose Dagster when you need a modern, asset-oriented approach to data orchestration with strong support for local development, testing, and ML integration.
 \item Choose Apache Airflow for complex batch workflow orchestration, especially when you need extensive integrations with other tools and services, or when you have a large team familiar with the Apache ecosystem.
 \item Choose Apache NiFi for real-time data ingestion and processing scenarios, or when you need a visual tool for designing and monitoring continuous dataflows.
\end{itemize} 
In some cases, these tools can be used together to create a comprehensive data engineering solution. 

\newpage
\subsection{Key Knowledge Representation Formalisms in AI}\label{app:krr}

\begin{figure}[!htbp]
    \centering
    \includegraphics[width=1.01\textwidth]{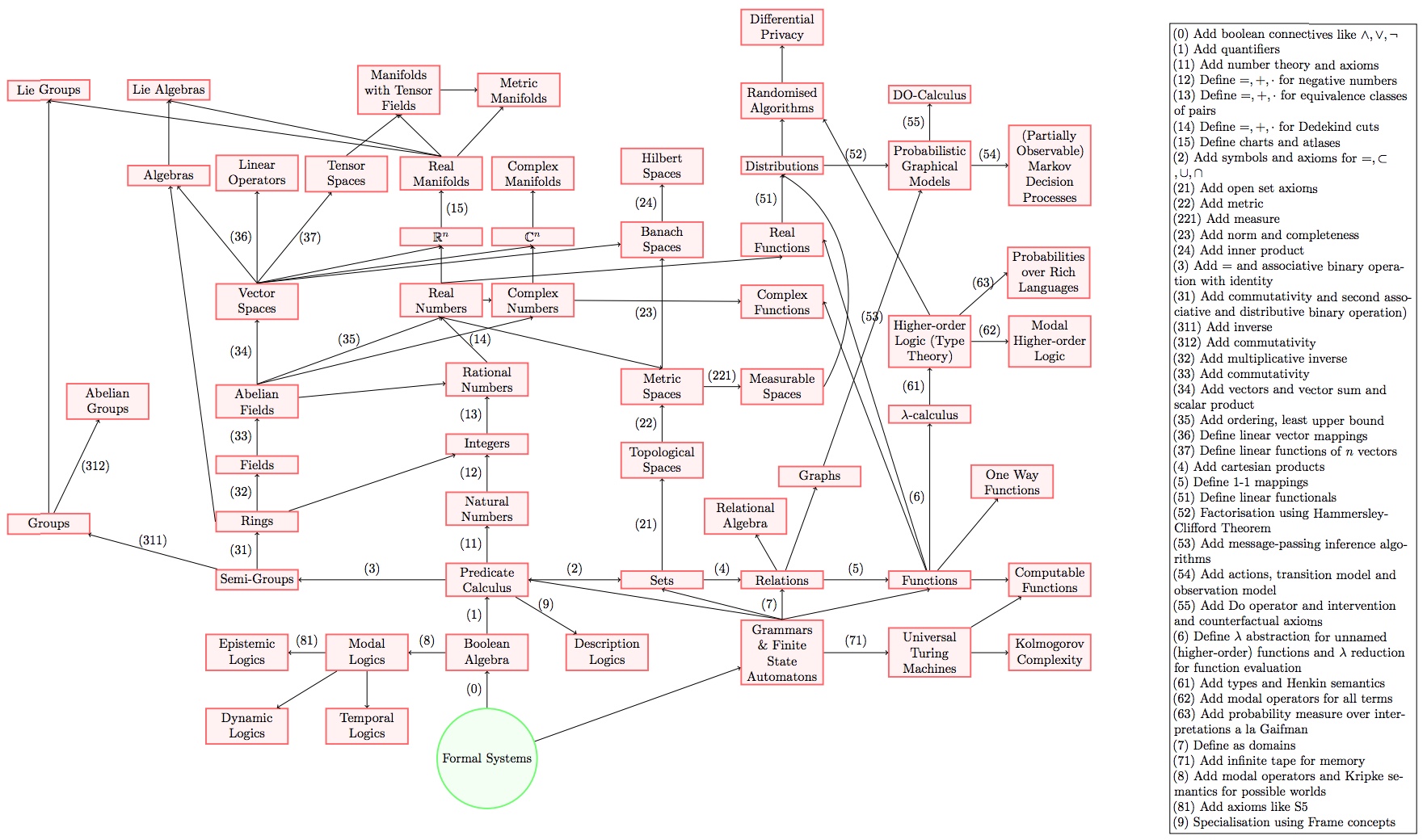}
    \caption{Major knowledge representation formalisms in AI}
    \label{fig:KRR}
\end{figure}

Figure~\ref{fig:KRR} shows a map of mathematical structures that are useful for thinking about knowledge representation and reasoning (KRR) issues in AI and ML. It is built on top of the diagram in \cite{tegmark1998theory} and extended with our own understanding of historical and recent work across quite a few different fields of AI. 
As such, it is necessarily biased towards our own personal experience and taste. 

Essentially all the systems in Figure~\ref{fig:KRR} have a syntax, a semantics in the styles of Tarski or Kripke, and a Hilbert-style proof procedure. The expressiveness of the different systems are tightly connected: each arrow in Figure~\ref{fig:KRR} involves the addition of some new symbols and the axioms that provide their definitions and/or properties. Some boxes have multiple incoming arrows; these are systems constructed from the union of multiple sets of new symbols and axioms. Note also that the relationships represented by the arrows are, in general, transitive.

The progression from propositional logic (Boolean algebra) to first-order logic (predicate calculus), second-order logic (Natural numbers) and ultimately higher-order logic (Type theory) is well covered in standard logic textbooks.
In the diagram, we have also shown how higher-order logic can be extended to modal higher-order logic \cite{lloyd2007knowledge} and probabilistic higher-order logic \cite{hutter2013probabilities}; the two extensions are orthogonal and it is possible to construct probabilistic multi-modal higher-order logic, where the interpretations are Kripke structures and there is a probability measure over those interpretations. 
All these logics can be shown to be sound and complete using Henkin semantics \cite{farmer2008seven}.

In AI, formal logics can be used primarily in one of two ways: (i) as a formal language for agent designers to specify systems, including multi-agent systems, and prove properties about them; (ii) as a formal knowledge representation language used inside an agent for it to represent and reason about the world. Generally speaking, modal logics and higher-order logics are usually adopted for the former, and first-order logics, the latter. The application of modal logics to modelling and reasoning about games \cite{van2014logic} and multi-agent systems have been an especially fruitful area, as is the application of higher-order logic to model and prove properties of complex software and hardware systems \cite{nipkow2002isabelle}. A notable exception is \cite{lloyd2007knowledge}, which is designed to be used inside an agent for learning and reasoning.

The study of mathematical structures like manifolds and Hilbert spaces near the top of Figure~\ref{fig:KRR} have been instrumental in many advances in the theory and practice of statistical machine learning, giving us a good understanding of dual representations of optimisation problems, 
different types of iterative gradient-based optimisation algorithms, and guidance on when to use what algorithms on which representations. Many of these ideas and algorithms underlie the strong results we are getting from kernel methods \cite{scholkopf2018learning} and deep neural networks \cite{lecun2015deep}.

The abstract algebra structures shown on the left of Figure~\ref{fig:KRR} are historically studied in cryptography. Recent advances in lattice-based cryptography, especially algebraic number theory, have given rise to reasonably efficient homomorphic encryption schemes based on the Ring Learning with Error problem \cite{li2022tutorial}. The use of such encryption schemes inside AI/ML algorithms has resulted in significant advances in Privacy-Preserving Machine Learning. 

On the upper right of Figure~\ref{fig:KRR}, we have Probabilistic Graphical Models \cite{koller2009probabilistic} and the special case of Markov Decision Processes (MDPs). 
These structures 
are foundational to AI/ML 
and they have delivered many practical applications in a range of areas. In the agent context, the study of Hidden Markov Models (HMMs) yielded widely used tracking algorithms like Kalman Filters, 
Particle Filters, and their multi-sensor multi-target variations. 
MDPs are HMMs that are augmented with action nodes and reward observations, and the study of MDPs and partially observable MDPs lie at the heart of the design and implementation of intelligent AI agents that can reason and learn to act optimally to achieve long-term expected rewards \cite{sutton2018reinforcement, mnih2015human}.

Some of the most exciting AI work happen in the intersection of probabilistic graphical models (in particular POMDPs), logic (first-order and higher-order logics), causal inference (Do Calculus), and privacy technologies (homomorphic encryption and differential privacy). 
We believe a modern AI and ML platform needs to support all these knowledge representation and reasoning formalisms to give practitioners the best chance of getting good results from these modern technologies.

\subsection{Key Machine Learning Principles and Algorithms}\label{app:ml algorithms}

\begin{figure}[!htbp]
    \centering
    \includegraphics[width=1.01\textwidth]{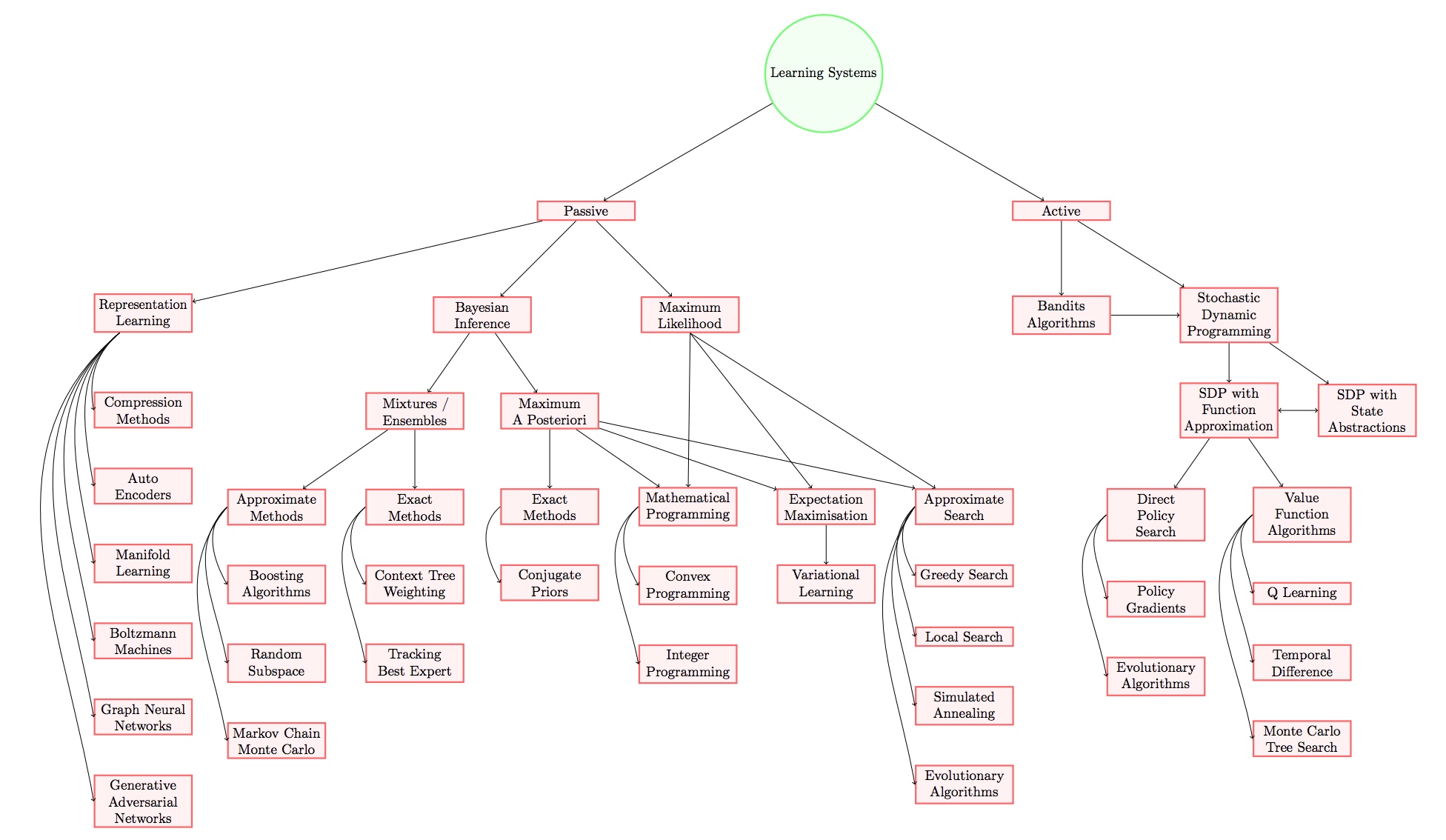}
    \caption{Major classes of machine learning algorithms}
    \label{fig:learning algorithms}
\end{figure}

Figure~\ref{fig:learning algorithms} shows the major classes of algorithms in Machine Learning, organised around the associated induction principles and learning theory. 

At the highest level, we can distinguish between Passive and Active Learning. In the passive case, the agent receives data chosen by the environment and its job is to learn a model to predict what is coming next. In the active case, in addition to receiving data from the environment, the agent can take actions and its goal is to learn to act 
to maximise some measure of long-term reward. In sufficiently rich domains, one can have nested passive and active settings, like in state-of-the-art large language models and reinforcement learning agents.

The most standard passive setting is the supervised learning setting, where the agent is given a set of labelled examples of the form $\{ (x_i,y_i) \}$ drawn from some unknown probability distribution, and the goal for the agent is to learn a model $f : X \rightarrow Y$ that, given arbitrary $x \in X$, can predict the corresponding $y \in Y$ with high probability of success. The labelled examples can come one at a time or in a large batch, and the underlying unknown probability distribution can stay the same or change over time. A variation of the supervised learning setting is the sequential prediction setting, where at every time step the agent has to predict the next observation given the history so far. 

In general terms, given the labelled examples $S := \{ (x_i,y_i) \}$ drawn from an unknown distribution $D$, a class $F$ of possible models, and a loss function $l(\cdot,\cdot)$, the agent’s goal is to find an $f \in F$ where the expected loss $l(f(x),y)$ is low for new previously unseen pairs $(x,y)$ drawn from $D$. 
There are different algorithmic strategies for learning the model. 
From Bayesian probability theory, the optimal solution is given by $f^*(x) = \sum_{f \in F} Pr(f | S) f(x)$, where $Pr(f | S) \propto Pr(S | f) Pr(f)$ is the posterior probability that $f$ is the true underlying model given we have seen $S$, and $Pr(f)$ is the prior probability of $f$, with less complex models given higher probabilities in accordance with Occam’s razor. The class of algorithms listed under Bayesian Inference -> Mixtures / Ensembles in Figure~\ref{fig:learning algorithms} are all algorithms that directly solve for $f^*$. In the rare cases where the model class $F$ has nice mathematical structures (like those that satisfy generalised distributive laws), the Bayesian optimal solution can be computed exactly and efficiently, and this strategy yields celebrated results like the Context Tree Weighting  \cite{willems1995context} and the Hedge family of algorithms \cite{cesa2006prediction}. In all other cases, an approximation 
is necessary.

The first approach, exemplified by the Boosting, Random Subspace and MCMC families of algorithms, is to approximate $f^*$ by averaging over only a subset of the models in $F$ that may have high posterior probabilities. Boosting \cite{freund1997decision} is a gradient-based approach to constructing an ensemble of weak-learners that can be shown to optimise a notion of margin on the training labelled examples. Random subspace methods like Bagging and Random Forests \cite{dietterich2000ensemble} combine random projection, bootstrapping, and majority voting ideas to build practical, low-variance and high accuracy ensemble models. In addition to those, there are also many Markov Chain Monte Carlo algorithms for estimating mixture models from data. 

The second approach to approximating $f^*$, named Maximum A Posteriori (MAP), is to find the model $f_{\it map} = \arg \max_{f\in F} Pr(S | f)Pr(f)$ in $F$ with the highest posterior probability. This optimisation can be solved exactly for distributions where the prior and the likelihood are conjugate distributions, in that their multiplication result in a new distribution with the same functional form. \footnote{See the list at \url{https://en.wikipedia.org/wiki/Conjugate_prior}} In cases where the MAP optimisation problem cannot be solved in closed form, we need to rely on the mathematical structure of the model class $F$ and the loss function to make progress. 
The field of Mathematical Programming / Optimisation is dedicated to the study of such problems, and a lot of work in Machine Learning is in designing how we can (re)formulate a learning-from-data problem as an efficiently solvable convex optimisation problem \cite{boyd2004convex}. 
If this cannot be done, one can resort to defining an upper / lower bound to the desired optimisation criteria and solve that easier problem instead, a technique generally referred to as variational inference \cite{blei2017variational}, which is closely related to expectation-maximisation. If even that cannot be done, e.g. when the model class $F$ is discrete, then one has to resort to approximate search algorithms that make use of local topology, or draw inspiration from biological processes (evolutionary search) or physical processes (simulated annealing).

The third and final approach to approximating $f^*$ is Maximum Likelihood, which is similar to the MAP approach except that we assume all models have equal prior probability, which means we seek the model that maximises the likelihood $f_{\it ml} = \arg \max_{f \in F} Pr(S | f)$. As for MAP estimation, the techniques of Mathematical Programming, Expectation Maximisation / Variational Inference, and Approximate search also apply to Maximum Likelihood estimation.

In Figure~\ref{fig:learning algorithms}, we have also shown a collection of techniques grouped under Representation Learning in the Passive setting. Compression methods include, among other things, clustering algorithms that are informed by the Minimum Description Length / Minimum Message Length principle, or algorithmic information-theoretic constructs like Normalized Compression Distance. Manifold Learning \cite{meilua2024manifold} include the different non-linear dimensionality reduction techniques that generalise Principal Component Analysis, including Isomap, Spectral Embedding, 
t-SNE, etc. 
Graph neural networks \cite{wu2020comprehensive}, and their predecessors like word2vec, are 
a class of techniques for computing embedding of words / tokens into numeric vectors that are then used in further downstream processing. These embeddings can usually capture some contextual semantics of words / tokens based on what other words / tokens tend to co-occur in their neighbourhood. Together with the Attention Mechanism \cite{vaswani2017attention}, embedding techniques like Auto-encoders and graph neural networks have been used very successfully as architectural components in Transformers that underlie a lot of the success behind large language models in recent years.

We now move on to Active learning. The simplest Active setup is the multi-armed bandit problem \cite{lattimore2020bandit}, where the agent has to come up with a strategy of choosing, at each time step, one of several possible actions, each of which has an unknown payoff. 
The solution requires optimally balancing exploration and exploitation, which is a highly non-trivial foundational problem in computer science. 
In the more general reinforcement learning setup, the agent can be in many different underlying states, which may or may not be directly observable, and the agent's goal is to learn from interactions with the environment an action-selection policy that can achieve long-term accumulated rewards.
The key challenge with reinforcement learning is that the reward signals are usually sparse, so the agent has to solve the credit-assignment problem even when most actions in most states result in no feedback from the environment.
This optimisation problem can be formulated using the Bellman equation, and 
algorithms like Q-learning and Temporal Difference learning can solve the Bellman equation exactly for small state and action spaces \cite{sutton2018reinforcement}. For problems with large state, observation, and action spaces, we usually have to resort to state abstraction and function approximation techniques to aproximate the different components of the Bellman equation. 
The different choices can lead to different algorithmic strategies, which are shown in Figure~\ref{fig:learning algorithms}.

Most, if not all, of the above techniques are now routinely used in both industry ML practices as well as state-of-the-art generative AI models.
We believe a modern AI and ML platform needs to support all these techniques to give practitioners the best chance of getting good results from these modern ML technologies.


\end{document}